\documentclass[aps,prl,reprint,groupedaddress]{revtex4-1}

\usepackage{amssymb, amsfonts, amsmath}
\usepackage{graphicx}
\usepackage{xcolor}

\begin{document}


\title{Lindemann unjamming of emulsions}


\author{Rodrigo E. Guerra}
\affiliation{Department of Physics, Harvard University, Cambridge, MA, USA}


\date{\today}

\begin{abstract}

We study the bulk and shear elastic properties of barely-compressed, ``athermal" emulsions and find that the rigidity of the jammed solid fails at remarkably large critical osmotic pressures.
The minuscule yield strain and similarly small Brownian particle displacement of solid emulsions close to this transition suggests that this catastrophic failure corresponds to a plastic-entropic instability: 
the solid becomes too soft and weak to resist the thermal agitation of the droplets that compose it and fails.
We propose a modified Lindemann stability criterion to describe this transition and derive a scaling law for the critical osmotic pressure that agrees quantitatively with experimental observations.

\end{abstract}


\maketitle

The mechanical properties of emulsions are controlled by two seemingly irreconcilable energy scales:
Dilute emulsions like cream and vinaigrette are fluids with osmotic moduli proportional to the ratio of thermal energy, $k_B T$, to droplet volume, $4\pi R^3/3$, while compressed emulsions like mayonnaise are jammed solids composed of droplets that are pressed together into amorphous, elastic packings with elastic moduli proportional to the ratio of interfacial tension, $\sigma$,   to droplet size, $R$ \cite{mason_elasticity_1995,saint-jalmes_vanishing_1999}.
However, when the applied strain exceeds a critical threshold, $\gamma_y$, droplets slide past each other and the solid yields.
The magnitudes of the shear modulus, $G$, and $\gamma_y$ are determined by the strength of the contacts between abutting droplets, which decrease with decreasing osmotic pressure, $\Pi$:
Reducing $\Pi$ thus makes the solid softer and more fragile, and this direct link between $\Pi$ and $G$ makes it possible to vary the shear modulus of a compressed emulsion over several orders of magnitude \cite{princen_osmotic_1987,princen_rheology_1986,mason_elasticity_1995,hebraud_yielding_1997,mason_yielding_1996,saint-jalmes_vanishing_1999}.
A smooth cross-over between this jammed elasticity and an entropic, glass-like rigidity has been observed for pastes and emulsions composed of sub-micron particles, whose thermal energy density, $\frac{3\,k_B T}{4\pi R^3}$, is large enough to easily match the modulus of the jammed solid \cite{koumakis_direct_2012,scheffold_linear_2013,mason_elasticity_1995,ikeda_disentangling_2013};
however, the gap between thermal and interfacial energy scales grows rapidly with increasing droplet size.
For emulsions composed of micrometer-scale droplets---which include food emulsions like mayonnaise and most emulsions produced by mechanical agitation---$\sigma/R$ can be 10$^6$ to 10$^{10}$ times larger than $\frac{3\,k_B T}{4\pi R^3}$, and it is not clear how this enormous energy gap is bridged.

Here, we study the pressure dependent shear and osmotic elasticity of barely-compressed, ``athermal" emulsions using Diffusing Wave Spectroscopy (DWS) microrheology and high-resolution magnetic resonance imaging.
We show that the shear rigidity of the jammed solid fails catastrophically and that its osmotic modulus declines rapidly below surprisingly large critical osmotic pressures, $\Pi^*\!\sim\!10^5\cdot\frac{3\,k_B T}{4\pi R^3}$, but minuscule droplet displacement amplitudes, $\sqrt{\langle\Delta r^2\rangle}/2R\sim$\,0.001.
We further find that this normalized droplet displacement amplitude coincides with the yield strain, $\gamma_y$, of an emulsion prepared close to its transition.
We propose a modified Lindemann stability criterion \cite{lindemann_uber_1910,gilvarry_lindemann_1956} that bridges these disparate energy scales and derive a critical scaling law for $\Pi^*$ that agrees quantitatively with the point where shear moduli determined microrheologically vanish and where osmotic moduli determined from magnetic resonance densitometry rapidly decline.
This instability is unlike anything seen or previously expected for three-dimensional solids, but should be common to a wide variety of soft materials that become softer and weaker at smaller osmotic pressures~\cite{menut_does_2011,saint-jalmes_vanishing_1999,koumakis_direct_2012}.

To obtain samples with well-known values of $\Pi$ we prepare a tall column of sedimented emulsion where the buoyant weight of the droplets themselves establishes a well-defined osmotic pressure gradient.
We prepare an emulsion composed of nearly-monodisperse, 7.2\,$\mu$m diameter droplets of anisole and polystyrene dispersed in a 2\,mM solution of sodium dodecylbenzenesulfonate in water.
The droplets are slightly denser than the surrounding water and we load enough of them into a rectangular glass tube to form a 20\,cm tall sediment.
The sample is maintained at a constant temperature, $T\!=$\,31.5$^\circ$C, and the sediment slowly consolidates and reaches mechanical equilibrium when the weight of every droplet is supported by the material beneath it~\cite{piazza_equilibrium_1993, biot_general_1941}.
Because the interface energy density of this emulsion, $\sigma/R\!=$\,1400\,Pa, is $\sim$10$^8$ times larger than its thermal energy density, $\frac{3\,k_B T}{4\pi R^3}\!=$\,16\,$\mu$Pa, emulsions like this are commonly referred to as ``athermal".

\begin{figure}
\includegraphics[width=8.8cm]{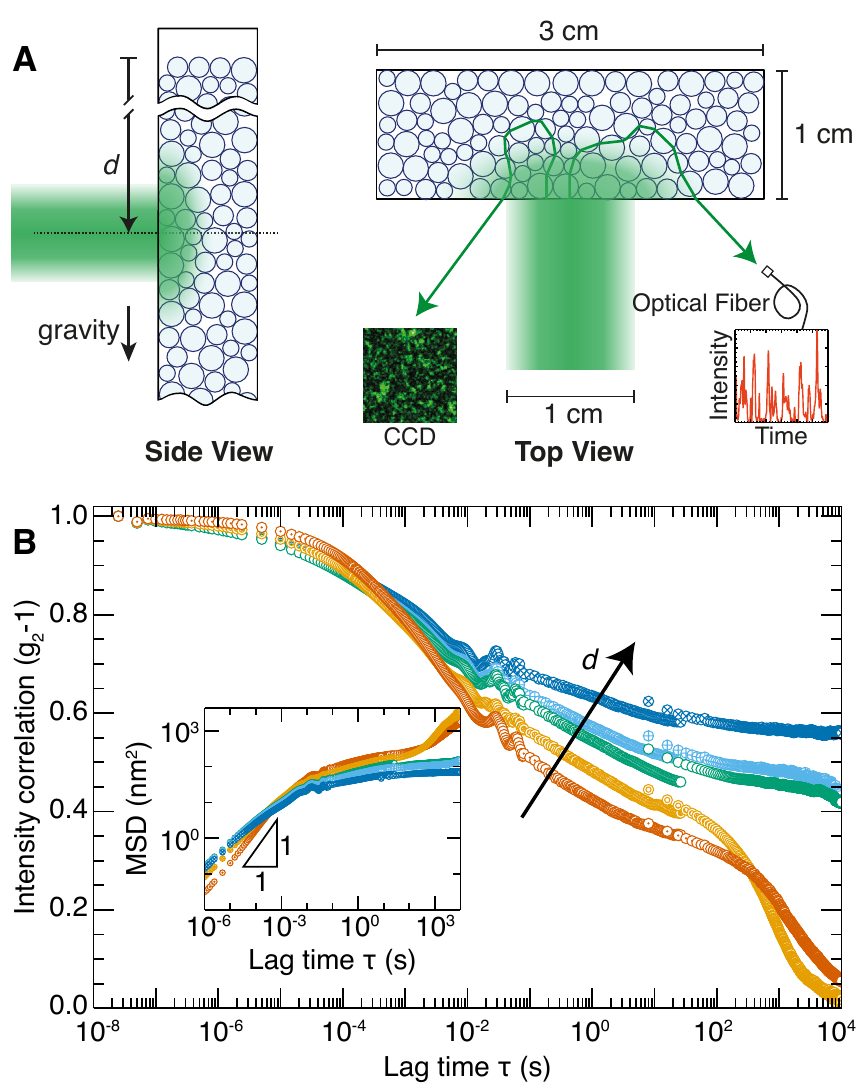}
\caption{\label{fig:DWScorr}
Diffusive light scattering from sedimented emulsion.
(A) 7.2\,$\mu$m diameter droplets are sealed in a thermostatted glass tube, where they settle and consolidate for several months.
The vertical position of the tube is adjusted so that the laser illuminates the sediment at a specific vertical distance below the top, $d$. A camera and optical fiber collect cross-polarized, backscattered light.
(B) Scattered light intensity autocorrelations, $g_2(\tau)$-1, measured at distances $d$\,=
0.5\,cm (\textcolor[RGB]{213, 94, 0}{{\raisebox{0.1em}{\fontsize{8}{12}\selectfont $\bigodot$}}}),
1.3\,cm (\textcolor[RGB]{230,159,0}{$\circledcirc$}),
4.0\,cm (\textcolor[RGB]{0,158,115}{{\raisebox{-0.2em}{\fontsize{20}{22}\selectfont $\circ$}}}),
4.9\,cm (\textcolor[RGB]{86,180,233}{{\raisebox{0.1em}{\fontsize{8}{12}\selectfont $\bigoplus$}}}),
and 8.5\,cm (\textcolor[RGB]{0,114,178}{{\raisebox{0.1em}{\fontsize{8}{12}\selectfont $\bigotimes$}}})
below the top of a sedimented emulsion held at 31.5$^{\circ}$C show clear separation between solid-like and fluid-like behaviors.
(Inset) Droplets closer to the bottom of the sediment reach a stable plateau MSD, while droplets closer to the top are slowed by the crowding of their neighbors but continue to move.}
\end{figure}

Even for these emulsions, thermal motion drives fluctuations in droplet positions that can be measured using dynamic light scattering and  Diffusing Wave Spectroscopy (DWS), which discern very small droplet motions interferometrically \cite{vera_scattering_2001,van_rossum_multiple_1999,lenke_coherent_2002,crassous_diffusive_2007,weitz_diffusing-wave_1993,mackintosh_diffusing-wave_1989,erpelding_diffusive_2008, mason_diffusing-wave-spectroscopy_1997}.
We illuminate the sediment at a prescribed vertical distance from the top, $d$, with a 1\,cm diameter, linearly-polarized laser beam, and collect cross-polarized, backscattered light using a camera and a split single-mode fiber connected to avalanche photodiodes (Fig.~\ref{fig:DWScorr}A).
We then autocorrelate the light intensities recorded by the camera and by the photodiodes, $I(t)$, and combine  them to compute $g_2(\tau)\! =\! \frac{\langle I(t+\tau)I(t)\rangle}{\langle I(t)\rangle\langle I(t+\tau)\rangle}$ for lag-times, $\tau$, spanning a combined twelve orders of magnitude~\cite{viasnoff_multispeckle_2002}.
The value of $g_2(\tau\!\to\!0)$\,-\,1 is normalized to one, and its decay for increasing $\tau$ is directly related to fluctuations in droplet position and shape.

We measure $g_2(\tau)$ at distances of 0.5\,cm, 1.3\,cm, 4.0\,cm, 4.9\,cm, and 8.5\,cm below the top of the sediment equilibrated at 31.5$^{\circ}$C.
The three correlation functions measured closest to the bottom of the sediment reach constant plateaus, consistent with solid-like elasticity, as shown in Fig.~\ref{fig:DWScorr}B.
Using DWS to relate $g_2(\tau)$ to the average mean squared displacement (MSD) of the illuminated droplet positions, $\langle\Delta r^2(\tau)\rangle$, we estimate that the magnitude of $\sqrt{\langle\Delta r^2(\tau\to\infty)\rangle}$ is less than 12\,nm for all three~\cite{mason_diffusing-wave-spectroscopy_1997,weitz_diffusing-wave_1993,mackintosh_diffusing-wave_1989,durian_multiple_1991}.
By contrast, the two correlation functions measured closer to the top of the sediment continue to decay, falling well below the noise floor of our instrument, and are clearly separate from the others.
Because of the limitations of DWS, we cannot determine how far the drops continue to move, but can conclude that the magnitudes of their long lag-time displacements are at least 4 times greater than the samples measured immediately below them.
The clear difference between these two behaviors is consistent with a sharp transition between a jammed solid and an entropic fluid or glass approximately 3\,cm below the top of the sediment.

\begin{figure*}
\includegraphics[width=18.3cm]{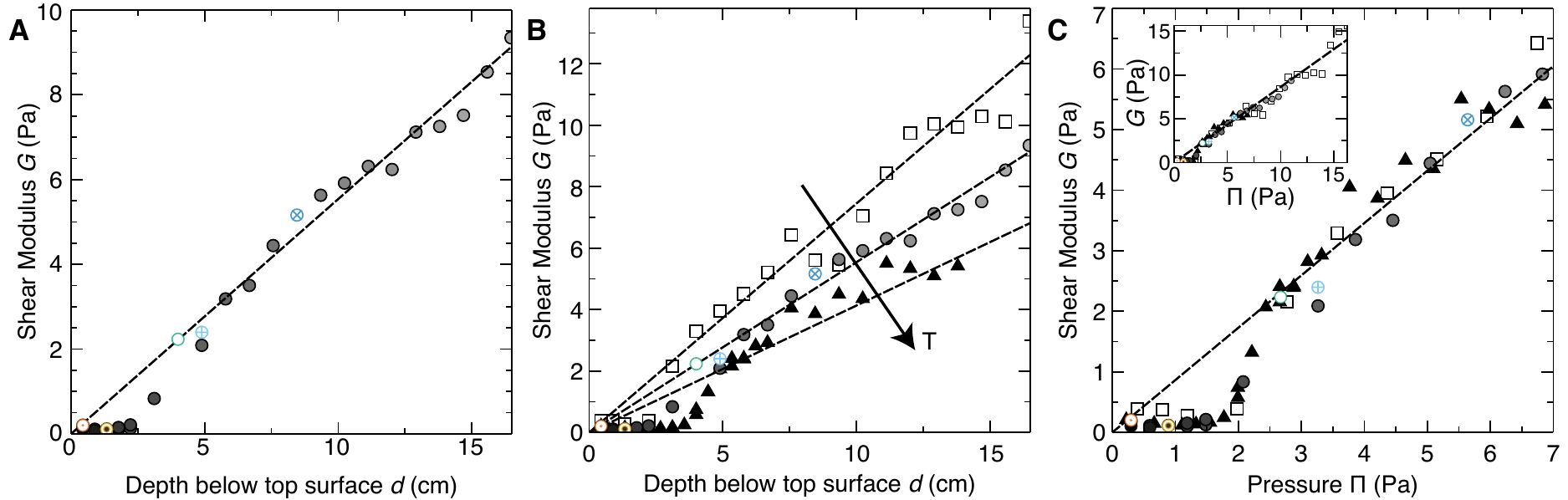}
\caption{\label{fig:DWSG}
Vertical depth and pressure dependence of sediment shear modulus.
(A) Vertical depth dependence of shear moduli, $G(d)$, inferred using DWS microrheology from the plateau values of $g_2(\tau)$ collected from a sediment held at 31.5$^{\circ}$C.
Colored symbols refer to moduli infered from data presented in Fig.~\ref{fig:DWScorr}B.
Dashed line is a guide for the eye.
(B) $G(d)$ measured for sediments held at 27$^{\circ}$C ($\square$), 31.5$^{\circ}$C (circles), and 34.9$^{\circ}$C ($\blacktriangle$).
(C) Osmotic pressure, $\Pi$, dependence of shear moduli shown in B.
Moduli measured at different temperatures collapse onto a single curve when $d$ is replaced by $\Pi$, and show that $G(\Pi)\!\sim\!\Pi$ when $\Pi\gtrsim2.5$\,Pa, but tends towards zero below it.
(Inset) Expanded range of pressures and shear moduli.}
\end{figure*}

To explore the effect of this transition on the shear modulus we use the plateau value of the MSD determined from the DWS measurement to calculate the shear modulus using microrheology, $G\!=\!\frac{k_B\,T}{\pi\,R\,\langle \Delta r^2(\tau\to\infty)\rangle}$~\cite{lin_force-displacement_2005}, and measure $G$ as a function of sample depth.
Near the top of the sample, where the emulsion is least compressed, $G\!=$\,0.
There is a sharp rise in modulus at the transition, and then $G$ increases linearly with depth, as shown in Fig.~\ref{fig:DWSG}A.
To expand the range of the data, we also make measurements of the sample after equilibriating it at 27.0$^{\circ}$C and 34.9$^{\circ}$C.
Changing temperature changes the buoyancy mismatch between the water and the oil, leading to a different height dependence of the osmotic pressure.
The resultant data show the same trend, with $G\!=$\,0 near the top, where the emulsion is least compressed, a sharp increase at the transition, followed by a linear increase in $G$ with height, as shown in Fig.~\ref{fig:DWSG}B.
To compare the three sets of measurements we estimate $\Pi(d)$ for each temperature and $d$ by measuring the density difference between the oil and water, $\delta\rho$, and assuming that the volume fraction near the bottom of the pile is not much larger than that near the top: $\Pi(d)\approx g\,\delta\rho(T)\,\phi_c\,d$, where $g$ is the gravitational acceleration and $\phi_c$ is the jamming or random close packing volume fraction (see Supplementary Information).
Replacing measurement depth with estimated pressure causes all the data collapse onto a single curve.
The data show that $G\approx\Pi$ for emulsions in the jammed state, but exhibit a sharp transition to a fluid state at $\Pi^*\!\approx$\,2.5\,Pa, as shown in Fig.~\ref{fig:DWSG}C.
The strict proportionality between $G$ and $\Pi$ is fundamentally incompatible with results from simulation \cite{goodrich_scaling_2016} but consistent with previous experimental data \cite{princen_osmotic_1987,princen_rheology_1986,saint-jalmes_vanishing_1999,mason_elasticity_1995}.

Surprisingly, though the value of $\Pi^*$ is much smaller than the interfacial energy density of the emulsion ($\sigma/R\!\approx$1400\,Pa), it is also several orders of magnitude larger than its thermal energy density ($\frac{3\,k_B T}{4\pi R^3}\!\approx$16\,$\mu$Pa), while the RMSD of droplets just above $\Pi^*$ is minuscule: barely above 10\,nm (Fig.~\ref{fig:DWScorr}B inset).
We thus postulate that this transition corresponds to a mechanical instability, and propose a heuristic, Lindemann-type stability criterion in which the critical confining pressure corresponds to the point where the normalized RMSD is equal to $\gamma_y$: \begin{equation}
\label{eq:P_equi}
    \sqrt{\langle \Delta r^2(\Pi^*)\rangle}/2R=\gamma_y(\Pi^*)
\end{equation}

In its simplest form, the Lindemann melting criterion \cite{lindemann_uber_1910,gilvarry_lindemann_1956} asserts that crystalline solids melt when the ratio of atomic RMSD to interatomic separation, $r$, exceeds a universal value, $\sqrt{\langle \Delta r^2\rangle}/r\!=\!\rho$.
And, though this assertion is not a thermodynamically accurate description of first-order melting transitions in equilibrium, a value of $\rho\!\approx\!0.1$ provides surprisingly good agreement with experimental measurements of this ratio for many crystalline solids.
However, this value of $\rho$ is two orders of magnitude larger than that inferred for our emulsion using DWS, and our proposed replacement of $\rho$ by $\gamma_y$ is rooted in the direct relationship between microscopic particle displacements and local strains \cite{bagi_stress_1996,bagi_analysis_2006}, as described in more detail in \S\,4 of the Supplementary Information.

To convert our modified criterion into an explicit equation for $\Pi^*$ we combine scaling relations for $\gamma_y$, $G$, and $\Pi$ determined by previous experimental studies far from the transition~\cite{princen_osmotic_1987,princen_rheology_1986, saint-jalmes_vanishing_1999,mason_elasticity_1995}:
\begin{equation}
\label{eq:scalings}
	\begin{aligned}
	\gamma_y(\phi)&=\frac{\phi-\phi_c}{2}\\
	G(\phi)&=\frac{\sigma}{R}\phi(\phi-\phi_c)\\
	\Pi &= G
	\end{aligned}
\end{equation}
with the equipartition relation to solve eq.~\ref{eq:P_equi} for $\Pi^*$, and find:
\begin{equation}
\label{eq:P_crit}
	\Pi^*=\left(\frac{k_B\,T\phi^2_c \sigma^2}{\pi\,R^5}\right)^\frac{1}{3}=\frac{\sigma}{R}\,\bar{T}^\frac{1}{3}\,(4\phi^2_c)^\frac{1}{3}\sim\frac{\sigma}{R}\,\bar{T}^\frac{1}{3}
\end{equation}
where $\bar{T}=\frac{k_B\,T}{4\pi\sigma R^2}$ is a reduced temperature.
We can similarly arrive at eq.~\ref{eq:P_crit} by equating, $E_y$, the work required to yield a microscopic volume, $V_0$, and $k_BT$:
\begin{equation}
    E_y=\frac{1}{2}V_0\,G\,\gamma^2_y=k_BT
\end{equation}
assuming a microscopic activation volume, $V_0=6\cdot\frac{4\pi R^3}{3}$, that coincides with the activation volume of shear transformation zones in metallic and colloidal glasses~\cite{schall_structural_2007,heggen_creation_2004}.
For this emulsion $\sigma/R\!=$\,1400\,Pa and $\bar{T}\!\approx$\,5\,$\cdot$\,10$^{-9}$, and eq.~\ref{eq:P_crit} evaluates to $\Pi^*\!\approx$\,2.8\,Pa: in remarkably close agreement with our measurements.

\begin{figure}
\includegraphics[width=8.8cm]{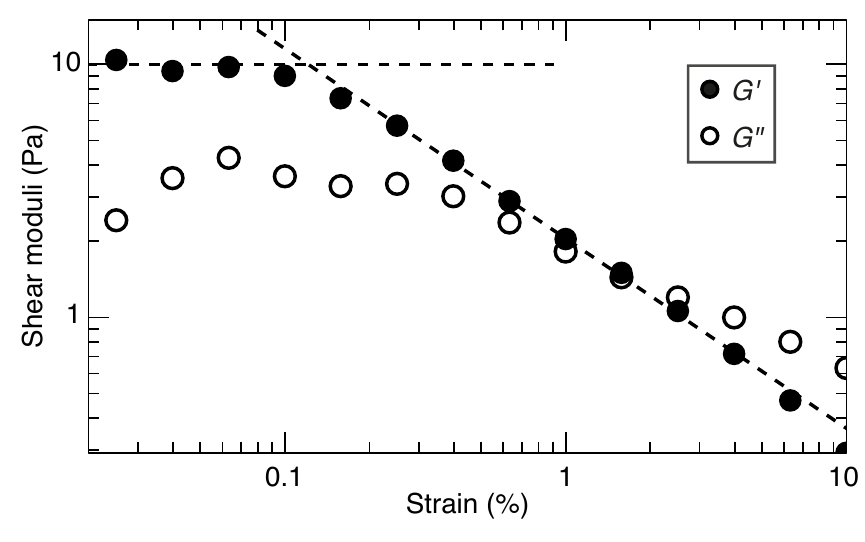}
\caption{\label{fig:binrheo}
Oscillatory rheology of weakly jammed emulsion.
Strain amplitude dependence of the elastic ($G'$, {\raisebox{-0.2em}{\fontsize{22}{24}\selectfont $\bullet$}}) and viscous ($G''$, {\raisebox{0.05em}{\fontsize{9}{12}\selectfont $\bigcirc$}}) shear moduli of a homogeneous emulsion measured in a mechanical rheometer at 0.005\,Hz.
The yield strain, $\gamma_y$, of this soft and fragile solid can be estimated as $\approx$\,0.1\% by the intersection of the dashed lines fit to the linear elastic and shear thinning regimes.}
\end{figure}

The value of $\gamma_y(\Pi^*)$ predicted from the empirical scaling relations in eq.~\ref{eq:scalings}, $\gamma_y(\Pi^*)\!\approx\!\frac{R\,\Pi}{2\,\sigma\phi_c}\!\approx$\,0.0015, agrees remarkably well with $\sqrt{\langle \Delta r^2(\Pi^*)\rangle}/2R\approx$\,0.0017, but is nevertheless strikingly small.
To test whether such a small value is valid for these emulsions, we prepare a reference sample that is close to the transition but is strong enough to measure with a conventional rheometer, and measure its yield strain.
We adjust the concentration of this emulsion by gentle centrifugation to obtain a linear shear modulus of 10\,Pa and use a double-Couette cell oscillating at 0.005\,Hz to measure the in phase and out of phase components of the shear modulus, $G'$ and $G''$ respectively, as a function of maximum strain, $\gamma$.
The elastic modulus, $G'$, is independent of strain at low $\gamma$, but begins to decay at $\gamma_y\approx0.001$, ultimately decreasing below $G''$ at larger $\gamma$, as shown in Fig.~\ref{fig:binrheo}.
Thus, these very low values of yield strain are indeed observed for these barely-compressed samples.

\begin{figure}
\includegraphics[width=8.8cm]{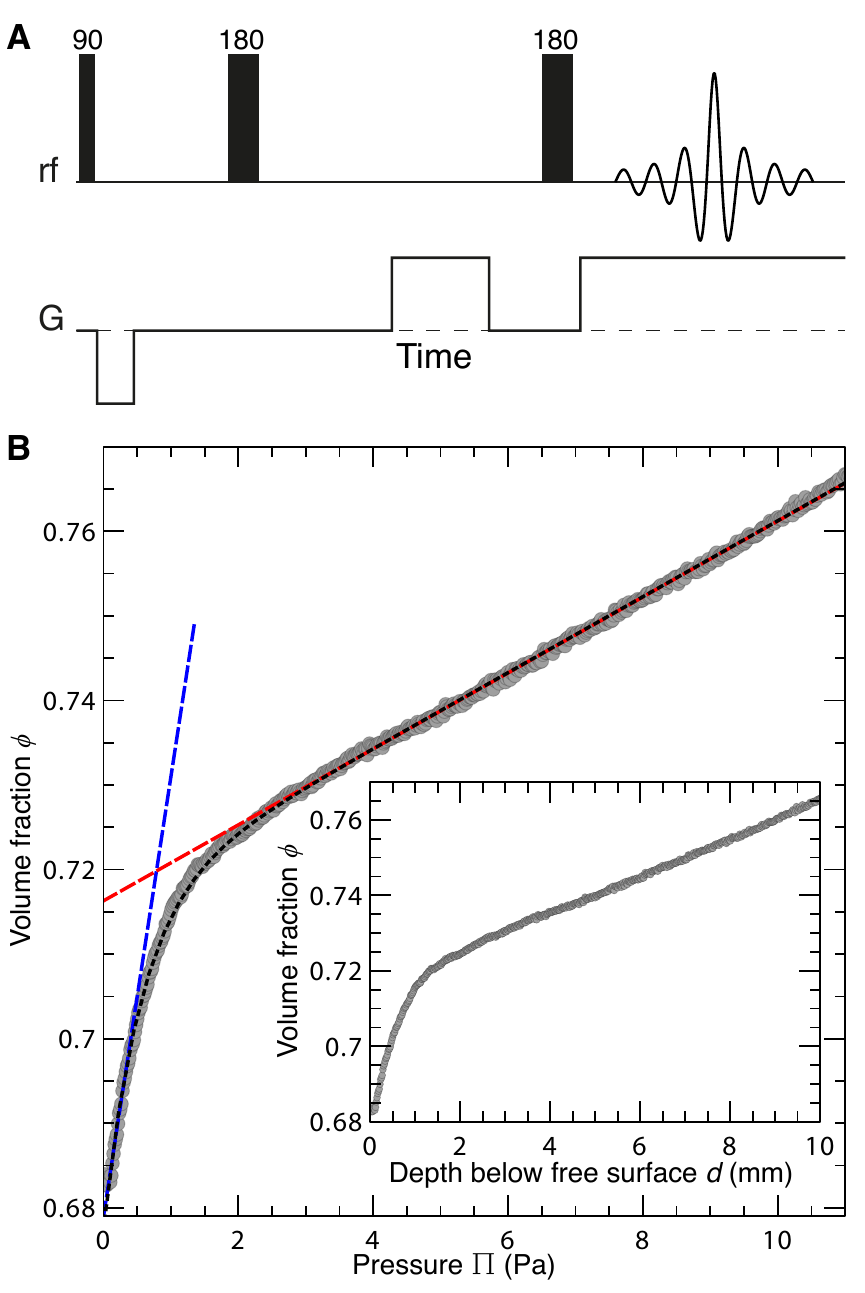}
\caption{\label{fig:NMR}
Magnetic densitometry of a sedimented emulsion.
The vertical depth dependence of droplet volume fraction, $\phi(d)$, of a sediment composed of oil droplets in D$_2$O measured with a NMR spectrometer programmed to act as a high-resolution densitometer.
(A) Simplified form of the spin- and gradient-echo pulse sequence used for proton imaging.
(B) Osmotic pressure dependence of oil volume fraction, $\phi(\Pi)$, computed from $\phi(d)$ (see text).
The dashed red line is proportional to $\frac{\Pi R}{\phi_c\,\sigma}$.
The dashed blue line is a linear fit to $\phi(\Pi)$ for small pressures.
The dashed black curve is a fit to eq.~\ref{eq:rhopi}.
(Inset) Direct measurement of $\phi(d)$.}
\end{figure}

Finally, to investigate how this mechanical instability affects the bulk modulus of ``athermal" emulsions we program a high-field nuclear magnetic resonance (NMR) spectrometer to serve as a magnetic resonance imaging densitometer, and prepare an emulsion formulated to provide precise, absolute measurements of $\phi(d)$.
We prepare monodisperse, 13.2\,$\mu$m diameter droplets of a mixture of silicone oil and tetrachloroethylene dispersed in a 2\,mM solution of sodium dodecylbenzenesulfonate in D$_2$O, load these droplets into an NMR tube, and wait several weeks for the sediment to consolidate.
A simplified version of our combined spin- and gradient-echo sequence~\cite{callaghan_principles_1993,torrey_bloch_1956} is shown in Fig.~\ref{fig:NMR}A.
We use a similar sequence to simultaneously measure the concentration of D$_2$O, which provides an absolute volume fraction reference.
Further details and calibrations are available in the Supplementary Information.

The volume fraction measured closer to the bottom of the sediment increases slowly and linearly, but its slope increases rapidly near the top (Fig.~\ref{fig:NMR}B, inset).
We compute $\phi(\Pi)$ by combining $\phi(d)$ with the integrated gravitational stress, $\Pi(d)\! =\! g\,\delta\rho\,\int_{0}^d\phi(z)\,\mathrm{d}z$, where $\delta\rho$\,=\,150\,kg/m$^3$ is the is the buoyant density of the droplets, and extract the interfacial energy density, $\sigma/R\!=$\,330\,Pa, by fitting $\phi(\Pi\!>$\,4\,Pa) to $\phi(\Pi)=\phi_c+\frac{\Pi R}{\sigma\,\phi_c}$~\cite{mason_elasticity_1995}.
A similar, though much steeper straight line also fits the data near the top of the sediment, and the extrapolated lines cross at $\Pi\!\approx$\,0.78\,Pa.
A tilted exponential interpolates both regimes, providing a good fit to all the data:
\begin{equation}
	\label{eq:rhopi}
	\phi(\Pi)=\phi_0+\frac{\Pi R}{\sigma\,\phi_0}-\delta\phi\,\exp{\left(-\frac{\Pi}{\Pi^*_\phi}\right)}
\end{equation}
with best-fit parameters $\phi_0\!=$\,0.716, $\sigma/R\!=$\,311\,Pa, $\delta\phi\!=$\,0.039, and $\Pi^*_\phi\!=$\,0.56\,Pa (Fig.~\ref{fig:NMR}B).
We use this value of $\sigma/R$ and $\bar{T}\!\approx$\,3.4\,$\cdot$\,10$^{-9}$ to compute the value of $\Pi^*$ predicted by eq.~\ref{eq:P_crit} and obtain $\Pi^*\! \approx$\,0.6\,Pa: again $\sim$\,10$^5$ times larger than $\frac{3\,k_B T}{4\pi R^3}$, but effectively identical to $\Pi^*_\phi$.

These results demonstrate the critical importance of thermal fluctuations and plasticity to the transition between jammed and entropic behaviors of soft solids.
The growing softness of the emulsion near this transition amplifies the displacements of thermally agitated droplets, and its fragility lets these displacements continually yield and restructure the emulsion.
Based on this intuitive picture we propose a simple stability criterion that 
accurately identifies the entropic-jammed boundary found from light scattering measurements, and points to a region of the equation of state where the bulk modulus of the emulsion rapidly decreases.
The combination of an abrupt drop in shear rigidity with a smoothly decreasing density sets this unjamming transition apart from conventional first-order melting or glass transitions.
However, given that a vanishing rigidity at low pressures is a common property of jammed materials, such an elastic instability may be an inescapable consequence and a universal property of such soft solids.

\begin{acknowledgments}
We would like to thank F. Spaepen, T. E. Kodger, and P. M. Chaikin for helpful conversations.
Magnetic resonance densitometry was performed at the Harvard Magnetic Resonance Facility.
We would like to thank Dr. Shaw Huang for his assistance with the implementation of this measurement.
This work was partially supported by the Harvard MRSEC (DMR-1420570).
\end{acknowledgments}


\begin{thebibliography}{32}%
\makeatletter
\providecommand \@ifxundefined [1]{%
 \@ifx{#1\undefined}
}%
\providecommand \@ifnum [1]{%
 \ifnum #1\expandafter \@firstoftwo
 \else \expandafter \@secondoftwo
 \fi
}%
\providecommand \@ifx [1]{%
 \ifx #1\expandafter \@firstoftwo
 \else \expandafter \@secondoftwo
 \fi
}%
\providecommand \natexlab [1]{#1}%
\providecommand \enquote  [1]{``#1''}%
\providecommand \bibnamefont  [1]{#1}%
\providecommand \bibfnamefont [1]{#1}%
\providecommand \citenamefont [1]{#1}%
\providecommand \href@noop [0]{\@secondoftwo}%
\providecommand \href [0]{\begingroup \@sanitize@url \@href}%
\providecommand \@href[1]{\@@startlink{#1}\@@href}%
\providecommand \@@href[1]{\endgroup#1\@@endlink}%
\providecommand \@sanitize@url [0]{\catcode `\\12\catcode `\$12\catcode
  `\&12\catcode `\#12\catcode `\^12\catcode `\_12\catcode `\%12\relax}%
\providecommand \@@startlink[1]{}%
\providecommand \@@endlink[0]{}%
\providecommand \url  [0]{\begingroup\@sanitize@url \@url }%
\providecommand \@url [1]{\endgroup\@href {#1}{\urlprefix }}%
\providecommand \urlprefix  [0]{URL }%
\providecommand \Eprint [0]{\href }%
\providecommand \doibase [0]{https://doi.org/}%
\providecommand \selectlanguage [0]{\@gobble}%
\providecommand \bibinfo  [0]{\@secondoftwo}%
\providecommand \bibfield  [0]{\@secondoftwo}%
\providecommand \translation [1]{[#1]}%
\providecommand \BibitemOpen [0]{}%
\providecommand \bibitemStop [0]{}%
\providecommand \bibitemNoStop [0]{.\EOS\space}%
\providecommand \EOS [0]{\spacefactor3000\relax}%
\providecommand \BibitemShut  [1]{\csname bibitem#1\endcsname}%
\let\auto@bib@innerbib\@empty
\bibitem [{\citenamefont {Mason}\ \emph {et~al.}(1995)\citenamefont {Mason},
  \citenamefont {Bibette},\ and\ \citenamefont
  {Weitz}}]{mason_elasticity_1995}%
  \BibitemOpen
  \bibfield  {author} {\bibinfo {author} {\bibfnamefont {T.~G.}\ \bibnamefont
  {Mason}}, \bibinfo {author} {\bibfnamefont {J.}~\bibnamefont {Bibette}},\
  and\ \bibinfo {author} {\bibfnamefont {D.~A.}\ \bibnamefont {Weitz}},\
  }\bibfield  {title} {\bibinfo {title} {Elasticity of {Compressed}
  {Emulsions}},\ }\href {https://doi.org/10.1103/PhysRevLett.75.2051}
  {\bibfield  {journal} {\bibinfo  {journal} {Physical Review Letters}\
  }\textbf {\bibinfo {volume} {75}},\ \bibinfo {pages} {2051} (\bibinfo {year}
  {1995})}\BibitemShut {NoStop}%
\bibitem [{\citenamefont {Saint-Jalmes}\ and\ \citenamefont
  {Durian}(1999)}]{saint-jalmes_vanishing_1999}%
  \BibitemOpen
  \bibfield  {author} {\bibinfo {author} {\bibfnamefont {A.}~\bibnamefont
  {Saint-Jalmes}}\ and\ \bibinfo {author} {\bibfnamefont {D.~J.}\ \bibnamefont
  {Durian}},\ }\bibfield  {title} {\bibinfo {title} {Vanishing elasticity for
  wet foams: {Equivalence} with emulsions and role of polydispersity},\ }\href
  {https://doi.org/10.1122/1.551052} {\bibfield  {journal} {\bibinfo  {journal}
  {Journal of Rheology}\ }\textbf {\bibinfo {volume} {43}},\ \bibinfo {pages}
  {1411} (\bibinfo {year} {1999})}\BibitemShut {NoStop}%
\bibitem [{\citenamefont {Princen}\ and\ \citenamefont
  {Kiss}(1987)}]{princen_osmotic_1987}%
  \BibitemOpen
  \bibfield  {author} {\bibinfo {author} {\bibfnamefont {H.~M.}\ \bibnamefont
  {Princen}}\ and\ \bibinfo {author} {\bibfnamefont {A.~D.}\ \bibnamefont
  {Kiss}},\ }\bibfield  {title} {\bibinfo {title} {Osmotic pressure of foams
  and highly concentrated emulsions. 2. {Determination} from the variation in
  volume fraction with height in an equilibrated column},\ }\href
  {https://doi.org/10.1021/la00073a007} {\bibfield  {journal} {\bibinfo
  {journal} {Langmuir}\ }\textbf {\bibinfo {volume} {3}},\ \bibinfo {pages}
  {36} (\bibinfo {year} {1987})}\BibitemShut {NoStop}%
\bibitem [{\citenamefont {Princen}\ and\ \citenamefont
  {Kiss}(1986)}]{princen_rheology_1986}%
  \BibitemOpen
  \bibfield  {author} {\bibinfo {author} {\bibfnamefont {H.~M.}\ \bibnamefont
  {Princen}}\ and\ \bibinfo {author} {\bibfnamefont {A.~D.}\ \bibnamefont
  {Kiss}},\ }\bibfield  {title} {\bibinfo {title} {Rheology of foams and highly
  concentrated emulsions: {III.} {Static} shear modulus},\ }\href
  {http://www.sciencedirect.com/science/article/pii/0021979786901116}
  {\bibfield  {journal} {\bibinfo  {journal} {Journal of colloid and interface
  science}\ }\textbf {\bibinfo {volume} {112}},\ \bibinfo {pages} {427}
  (\bibinfo {year} {1986})}\BibitemShut {NoStop}%
\bibitem [{\citenamefont {H{\'e}braud}\ \emph {et~al.}(1997)\citenamefont
  {H{\'e}braud}, \citenamefont {Lequeux}, \citenamefont {M{\"u}nch},\ and\
  \citenamefont {Pine}}]{hebraud_yielding_1997}%
  \BibitemOpen
  \bibfield  {author} {\bibinfo {author} {\bibfnamefont {P.}~\bibnamefont
  {H{\'e}braud}}, \bibinfo {author} {\bibfnamefont {F.}~\bibnamefont
  {Lequeux}}, \bibinfo {author} {\bibfnamefont {J.-P.}\ \bibnamefont
  {M{\"u}nch}},\ and\ \bibinfo {author} {\bibfnamefont {D.~J.}\ \bibnamefont
  {Pine}},\ }\bibfield  {title} {\bibinfo {title} {Yielding and
  {Rearrangements} in {Disordered} {Emulsions}},\ }\href
  {https://doi.org/10.1103/PhysRevLett.78.4657} {\bibfield  {journal} {\bibinfo
   {journal} {Physical Review Letters}\ }\textbf {\bibinfo {volume} {78}},\
  \bibinfo {pages} {4657} (\bibinfo {year} {1997})}\BibitemShut {NoStop}%
\bibitem [{\citenamefont {Mason}\ \emph {et~al.}(1996)\citenamefont {Mason},
  \citenamefont {Bibette},\ and\ \citenamefont {Weitz}}]{mason_yielding_1996}%
  \BibitemOpen
  \bibfield  {author} {\bibinfo {author} {\bibfnamefont {T.~G.}\ \bibnamefont
  {Mason}}, \bibinfo {author} {\bibfnamefont {J.}~\bibnamefont {Bibette}},\
  and\ \bibinfo {author} {\bibfnamefont {D.~A.}\ \bibnamefont {Weitz}},\
  }\bibfield  {title} {\bibinfo {title} {Yielding and flow of monodisperse
  emulsions},\ }\href
  {http://www.sciencedirect.com/science/article/pii/S0021979796902350}
  {\bibfield  {journal} {\bibinfo  {journal} {Journal of Colloid and Interface
  Science}\ }\textbf {\bibinfo {volume} {179}},\ \bibinfo {pages} {439}
  (\bibinfo {year} {1996})}\BibitemShut {NoStop}%
\bibitem [{\citenamefont {Koumakis}\ \emph {et~al.}(2012)\citenamefont
  {Koumakis}, \citenamefont {Pamvouxoglou},\ and\ \citenamefont
  {Poulos}}]{koumakis_direct_2012}%
  \BibitemOpen
  \bibfield  {author} {\bibinfo {author} {\bibfnamefont {N.}~\bibnamefont
  {Koumakis}}, \bibinfo {author} {\bibfnamefont {A.}~\bibnamefont
  {Pamvouxoglou}},\ and\ \bibinfo {author} {\bibfnamefont {G.}~\bibnamefont
  {Poulos}, \bibfnamefont {Andreas S. and~Petekidis}},\ }\bibfield  {title}
  {\bibinfo {title} {Direct comparison of the rheology of model hard and soft
  particle glasses},\ }\href {https://doi.org/10.1039/C2SM07113D} {\bibfield
  {journal} {\bibinfo  {journal} {Soft Matter}\ }\textbf {\bibinfo {volume}
  {8}},\ \bibinfo {pages} {4271} (\bibinfo {year} {2012})}\BibitemShut
  {NoStop}%
\bibitem [{\citenamefont {Scheffold}\ \emph {et~al.}(2013)\citenamefont
  {Scheffold}, \citenamefont {Cardinaux},\ and\ \citenamefont
  {Mason}}]{scheffold_linear_2013}%
  \BibitemOpen
  \bibfield  {author} {\bibinfo {author} {\bibfnamefont {F.}~\bibnamefont
  {Scheffold}}, \bibinfo {author} {\bibfnamefont {F.}~\bibnamefont
  {Cardinaux}},\ and\ \bibinfo {author} {\bibfnamefont {T.~G.}\ \bibnamefont
  {Mason}},\ }\bibfield  {title} {\bibinfo {title} {Linear and nonlinear
  rheology of dense emulsions across the glass and the jamming regimes},\
  }\href {https://doi.org/10.1088/0953-8984/25/50/502101} {\bibfield  {journal}
  {\bibinfo  {journal} {Journal of Physics: Condensed Matter}\ }\textbf
  {\bibinfo {volume} {25}},\ \bibinfo {pages} {502101} (\bibinfo {year}
  {2013})},\ \bibinfo {note} {00006}\BibitemShut {NoStop}%
\bibitem [{\citenamefont {Ikeda}\ \emph {et~al.}(2013)\citenamefont {Ikeda},
  \citenamefont {Berthier},\ and\ \citenamefont
  {Sollich}}]{ikeda_disentangling_2013}%
  \BibitemOpen
  \bibfield  {author} {\bibinfo {author} {\bibfnamefont {A.}~\bibnamefont
  {Ikeda}}, \bibinfo {author} {\bibfnamefont {L.}~\bibnamefont {Berthier}},\
  and\ \bibinfo {author} {\bibfnamefont {P.}~\bibnamefont {Sollich}},\
  }\bibfield  {title} {\bibinfo {title} {Disentangling glass and jamming
  physics in the rheology of soft materials},\ }\href
  {https://doi.org/10.1039/c3sm50503k} {\bibfield  {journal} {\bibinfo
  {journal} {Soft Matter}\ }\textbf {\bibinfo {volume} {9}},\ \bibinfo {pages}
  {7669} (\bibinfo {year} {2013})}\BibitemShut {NoStop}%
\bibitem [{\citenamefont {Lindemann}(1910)}]{lindemann_uber_1910}%
  \BibitemOpen
  \bibfield  {author} {\bibinfo {author} {\bibfnamefont {V.~F.~A.}\
  \bibnamefont {Lindemann}},\ }\bibfield  {title} {\bibinfo {title} {{\"U}ber
  die {Berechnung} molekularer {Eigenfrequenzen}},\ }\href@noop {} {\bibfield
  {journal} {\bibinfo  {journal} {Physikalische Zeitschrift}\ }\textbf
  {\bibinfo {volume} {10}},\ \bibinfo {pages} {609} (\bibinfo {year}
  {1910})}\BibitemShut {NoStop}%
\bibitem [{\citenamefont {Gilvarry}(1956)}]{gilvarry_lindemann_1956}%
  \BibitemOpen
  \bibfield  {author} {\bibinfo {author} {\bibfnamefont {J.~J.}\ \bibnamefont
  {Gilvarry}},\ }\bibfield  {title} {\bibinfo {title} {The {Lindemann} and
  {Gr{\"u}neisen} {Laws}},\ }\href {https://doi.org/10.1103/PhysRev.102.308}
  {\bibfield  {journal} {\bibinfo  {journal} {Physical Review}\ }\textbf
  {\bibinfo {volume} {102}},\ \bibinfo {pages} {308} (\bibinfo {year}
  {1956})}\BibitemShut {NoStop}%
\bibitem [{\citenamefont {Menut}\ \emph {et~al.}(2011)\citenamefont {Menut},
  \citenamefont {Seiffert}, \citenamefont {Sprakel},\ and\ \citenamefont
  {Weitz}}]{menut_does_2011}%
  \BibitemOpen
  \bibfield  {author} {\bibinfo {author} {\bibfnamefont {P.}~\bibnamefont
  {Menut}}, \bibinfo {author} {\bibfnamefont {S.}~\bibnamefont {Seiffert}},
  \bibinfo {author} {\bibfnamefont {J.}~\bibnamefont {Sprakel}},\ and\ \bibinfo
  {author} {\bibfnamefont {D.~A.}\ \bibnamefont {Weitz}},\ }\bibfield  {title}
  {\bibinfo {title} {Does size matter? {Elasticity} of compressed suspensions
  of colloidal- and granular-scale microgels},\ }\href
  {https://doi.org/10.1039/C1SM06355C} {\bibfield  {journal} {\bibinfo
  {journal} {Soft Matter}\ }\textbf {\bibinfo {volume} {8}},\ \bibinfo {pages}
  {156} (\bibinfo {year} {2011})}\BibitemShut {NoStop}%
\bibitem [{\citenamefont {Piazza}\ \emph {et~al.}(1993)\citenamefont {Piazza},
  \citenamefont {Bellini},\ and\ \citenamefont
  {Degiorgio}}]{piazza_equilibrium_1993}%
  \BibitemOpen
  \bibfield  {author} {\bibinfo {author} {\bibfnamefont {R.}~\bibnamefont
  {Piazza}}, \bibinfo {author} {\bibfnamefont {T.}~\bibnamefont {Bellini}},\
  and\ \bibinfo {author} {\bibfnamefont {V.}~\bibnamefont {Degiorgio}},\
  }\bibfield  {title} {\bibinfo {title} {Equilibrium sedimentation profiles of
  screened charged colloids: {A} test of the hard-sphere equation of state},\
  }\href {http://prl.aps.org/abstract/PRL/v71/i25/p4267_1} {\bibfield
  {journal} {\bibinfo  {journal} {Physical Review Letters}\ }\textbf {\bibinfo
  {volume} {71}},\ \bibinfo {pages} {4267} (\bibinfo {year}
  {1993})}\BibitemShut {NoStop}%
\bibitem [{\citenamefont {Biot}(1941)}]{biot_general_1941}%
  \BibitemOpen
  \bibfield  {author} {\bibinfo {author} {\bibfnamefont {M.~A.}\ \bibnamefont
  {Biot}},\ }\bibfield  {title} {\bibinfo {title} {General {Theory} of
  {Three}-{Dimensional} {Consolidation}},\ }\href
  {https://doi.org/10.1063/1.1712886} {\bibfield  {journal} {\bibinfo
  {journal} {Journal of Applied Physics}\ }\textbf {\bibinfo {volume} {12}},\
  \bibinfo {pages} {155} (\bibinfo {year} {1941})}\BibitemShut {NoStop}%
\bibitem [{\citenamefont {Vera}\ \emph {et~al.}(2001)\citenamefont {Vera},
  \citenamefont {Saint-Jalmes},\ and\ \citenamefont
  {Durian}}]{vera_scattering_2001}%
  \BibitemOpen
  \bibfield  {author} {\bibinfo {author} {\bibfnamefont {M.~U.}\ \bibnamefont
  {Vera}}, \bibinfo {author} {\bibfnamefont {A.}~\bibnamefont {Saint-Jalmes}},\
  and\ \bibinfo {author} {\bibfnamefont {D.~J.}\ \bibnamefont {Durian}},\
  }\bibfield  {title} {\bibinfo {title} {Scattering optics of foam},\ }\href
  {http://www.opticsinfobase.org/abstract.cfm?&id=65065} {\bibfield  {journal}
  {\bibinfo  {journal} {Applied Optics}\ }\textbf {\bibinfo {volume} {40}},\
  \bibinfo {pages} {4210} (\bibinfo {year} {2001})}\BibitemShut {NoStop}%
\bibitem [{\citenamefont {van Rossum}\ and\ \citenamefont
  {Nieuwenhuizen}(1999)}]{van_rossum_multiple_1999}%
  \BibitemOpen
  \bibfield  {author} {\bibinfo {author} {\bibfnamefont {M.~C.~W.}\
  \bibnamefont {van Rossum}}\ and\ \bibinfo {author} {\bibfnamefont {T.~M.}\
  \bibnamefont {Nieuwenhuizen}},\ }\bibfield  {title} {\bibinfo {title}
  {Multiple scattering of classical waves: microscopy, mesoscopy, and
  diffusion},\ }\href {https://doi.org/10.1103/RevModPhys.71.313} {\bibfield
  {journal} {\bibinfo  {journal} {Reviews of Modern Physics}\ }\textbf
  {\bibinfo {volume} {71}},\ \bibinfo {pages} {313} (\bibinfo {year}
  {1999})}\BibitemShut {NoStop}%
\bibitem [{\citenamefont {Lenke}\ \emph {et~al.}(2002)\citenamefont {Lenke},
  \citenamefont {Tweer},\ and\ \citenamefont {Maret}}]{lenke_coherent_2002}%
  \BibitemOpen
  \bibfield  {author} {\bibinfo {author} {\bibfnamefont {R.}~\bibnamefont
  {Lenke}}, \bibinfo {author} {\bibfnamefont {R.}~\bibnamefont {Tweer}},\ and\
  \bibinfo {author} {\bibfnamefont {G.}~\bibnamefont {Maret}},\ }\bibfield
  {title} {\bibinfo {title} {Coherent backscattering of turbid samples
  containing large {Mie} spheres},\ }\href
  {https://doi.org/10.1088/1464-4258/4/3/313} {\bibfield  {journal} {\bibinfo
  {journal} {Journal of Optics A: Pure and Applied Optics}\ }\textbf {\bibinfo
  {volume} {4}},\ \bibinfo {pages} {293} (\bibinfo {year} {2002})}\BibitemShut
  {NoStop}%
\bibitem [{\citenamefont {Crassous}(2007)}]{crassous_diffusive_2007}%
  \BibitemOpen
  \bibfield  {author} {\bibinfo {author} {\bibfnamefont {J.}~\bibnamefont
  {Crassous}},\ }\bibfield  {title} {\bibinfo {title} {Diffusive {Wave}
  {Spectroscopy} of a random close packing of spheres},\ }\href
  {https://doi.org/10.1140/epje/i2006-10079-y} {\bibfield  {journal} {\bibinfo
  {journal} {The European Physical Journal E}\ }\textbf {\bibinfo {volume}
  {23}},\ \bibinfo {pages} {145} (\bibinfo {year} {2007})}\BibitemShut
  {NoStop}%
\bibitem [{\citenamefont {Weitz}\ and\ \citenamefont
  {Pine}(1993)}]{weitz_diffusing-wave_1993}%
  \BibitemOpen
  \bibfield  {author} {\bibinfo {author} {\bibfnamefont {D.~A.}\ \bibnamefont
  {Weitz}}\ and\ \bibinfo {author} {\bibfnamefont {D.~J.}\ \bibnamefont
  {Pine}},\ }\bibfield  {title} {\bibinfo {title} {Diffusing-wave
  spectroscopy},\ }in\ \href@noop {} {\emph {\bibinfo {booktitle} {Dynamic
  light scattering: the method and some applications}}}\ (\bibinfo  {publisher}
  {Clarendon Press ; Oxford University Press},\ \bibinfo {address} {Oxford
  [England]; New York},\ \bibinfo {year} {1993})\ pp.\ \bibinfo {pages}
  {652--720}\BibitemShut {NoStop}%
\bibitem [{\citenamefont {MacKintosh}\ and\ \citenamefont
  {John}(1989)}]{mackintosh_diffusing-wave_1989}%
  \BibitemOpen
  \bibfield  {author} {\bibinfo {author} {\bibfnamefont {F.~C.}\ \bibnamefont
  {MacKintosh}}\ and\ \bibinfo {author} {\bibfnamefont {S.}~\bibnamefont
  {John}},\ }\bibfield  {title} {\bibinfo {title} {Diffusing-wave spectroscopy
  and multiple scattering of light in correlated random media},\ }\href
  {https://doi.org/10.1103/PhysRevB.40.2383} {\bibfield  {journal} {\bibinfo
  {journal} {Physical Review B}\ }\textbf {\bibinfo {volume} {40}},\ \bibinfo
  {pages} {2383} (\bibinfo {year} {1989})}\BibitemShut {NoStop}%
\bibitem [{\citenamefont {Erpelding}\ \emph {et~al.}(2008)\citenamefont
  {Erpelding}, \citenamefont {Amon},\ and\ \citenamefont
  {Crassous}}]{erpelding_diffusive_2008}%
  \BibitemOpen
  \bibfield  {author} {\bibinfo {author} {\bibfnamefont {M.}~\bibnamefont
  {Erpelding}}, \bibinfo {author} {\bibfnamefont {A.}~\bibnamefont {Amon}},\
  and\ \bibinfo {author} {\bibfnamefont {J.}~\bibnamefont {Crassous}},\
  }\bibfield  {title} {\bibinfo {title} {Diffusive wave spectroscopy applied to
  the spatially resolved deformation of a solid},\ }\href
  {https://doi.org/10.1103/PhysRevE.78.046104} {\bibfield  {journal} {\bibinfo
  {journal} {Physical Review E}\ }\textbf {\bibinfo {volume} {78}},\ \bibinfo
  {pages} {046104} (\bibinfo {year} {2008})}\BibitemShut {NoStop}%
\bibitem [{\citenamefont {Mason}\ \emph {et~al.}(1997)\citenamefont {Mason},
  \citenamefont {Gang},\ and\ \citenamefont
  {Weitz}}]{mason_diffusing-wave-spectroscopy_1997}%
  \BibitemOpen
  \bibfield  {author} {\bibinfo {author} {\bibfnamefont {T.~G.}\ \bibnamefont
  {Mason}}, \bibinfo {author} {\bibfnamefont {H.}~\bibnamefont {Gang}},\ and\
  \bibinfo {author} {\bibfnamefont {D.~A.}\ \bibnamefont {Weitz}},\ }\bibfield
  {title} {\bibinfo {title} {Diffusing-wave-spectroscopy measurements of
  viscoelasticity of complex fluids},\ }\href
  {https://doi.org/10.1364/JOSAA.14.000139} {\bibfield  {journal} {\bibinfo
  {journal} {Journal of the Optical Society of America A}\ }\textbf {\bibinfo
  {volume} {14}},\ \bibinfo {pages} {139} (\bibinfo {year} {1997})}\BibitemShut
  {NoStop}%
\bibitem [{\citenamefont {Viasnoff}\ \emph {et~al.}(2002)\citenamefont
  {Viasnoff}, \citenamefont {Lequeux},\ and\ \citenamefont
  {Pine}}]{viasnoff_multispeckle_2002}%
  \BibitemOpen
  \bibfield  {author} {\bibinfo {author} {\bibfnamefont {V.}~\bibnamefont
  {Viasnoff}}, \bibinfo {author} {\bibfnamefont {F.}~\bibnamefont {Lequeux}},\
  and\ \bibinfo {author} {\bibfnamefont {D.~J.}\ \bibnamefont {Pine}},\
  }\bibfield  {title} {\bibinfo {title} {Multispeckle diffusing-wave
  spectroscopy: {A} tool to study slow relaxation and time-dependent
  dynamics},\ }\href {https://doi.org/10.1063/1.1476699} {\bibfield  {journal}
  {\bibinfo  {journal} {Review of Scientific Instruments}\ }\textbf {\bibinfo
  {volume} {73}},\ \bibinfo {pages} {2336} (\bibinfo {year}
  {2002})}\BibitemShut {NoStop}%
\bibitem [{\citenamefont {Durian}\ \emph {et~al.}(1991)\citenamefont {Durian},
  \citenamefont {Weitz},\ and\ \citenamefont {Pine}}]{durian_multiple_1991}%
  \BibitemOpen
  \bibfield  {author} {\bibinfo {author} {\bibfnamefont {D.~J.}\ \bibnamefont
  {Durian}}, \bibinfo {author} {\bibfnamefont {D.~A.}\ \bibnamefont {Weitz}},\
  and\ \bibinfo {author} {\bibfnamefont {D.~J.}\ \bibnamefont {Pine}},\
  }\bibfield  {title} {\bibinfo {title} {Multiple {Light}-{Scattering} {Probes}
  of {Foam} {Structure} and {Dynamics}},\ }\href
  {http://www.jstor.org/stable/2875429} {\bibfield  {journal} {\bibinfo
  {journal} {Science}\ }\bibinfo {series} {New {Series}},\ \textbf {\bibinfo
  {volume} {252}},\ \bibinfo {pages} {686} (\bibinfo {year}
  {1991})}\BibitemShut {NoStop}%
\bibitem [{\citenamefont {Lin}\ \emph {et~al.}(2005)\citenamefont {Lin},
  \citenamefont {Langrana},\ and\ \citenamefont
  {Yurke}}]{lin_force-displacement_2005}%
  \BibitemOpen
  \bibfield  {author} {\bibinfo {author} {\bibfnamefont {D.~C.}\ \bibnamefont
  {Lin}}, \bibinfo {author} {\bibfnamefont {N.~A.}\ \bibnamefont {Langrana}},\
  and\ \bibinfo {author} {\bibfnamefont {B.}~\bibnamefont {Yurke}},\ }\bibfield
   {title} {\bibinfo {title} {Force-displacement relationships for spherical
  inclusions in finite elastic media},\ }\href
  {https://doi.org/10.1063/1.1847698} {\bibfield  {journal} {\bibinfo
  {journal} {Journal of Applied Physics}\ }\textbf {\bibinfo {volume} {97}},\
  \bibinfo {pages} {043510} (\bibinfo {year} {2005})}\BibitemShut {NoStop}%
\bibitem [{\citenamefont {Goodrich}\ \emph {et~al.}(2016)\citenamefont
  {Goodrich}, \citenamefont {Liu},\ and\ \citenamefont
  {Sethna}}]{goodrich_scaling_2016}%
  \BibitemOpen
  \bibfield  {author} {\bibinfo {author} {\bibfnamefont {C.~P.}\ \bibnamefont
  {Goodrich}}, \bibinfo {author} {\bibfnamefont {A.~J.}\ \bibnamefont {Liu}},\
  and\ \bibinfo {author} {\bibfnamefont {J.~P.}\ \bibnamefont {Sethna}},\
  }\bibfield  {title} {\bibinfo {title} {Scaling ansatz for the jamming
  transition},\ }\href {https://doi.org/10.1073/pnas.1601858113} {\bibfield
  {journal} {\bibinfo  {journal} {Proceedings of the National Academy of
  Sciences}\ }\textbf {\bibinfo {volume} {113}},\ \bibinfo {pages} {9745}
  (\bibinfo {year} {2016})}\BibitemShut {NoStop}%
\bibitem [{\citenamefont {Bagi}(1996)}]{bagi_stress_1996}%
  \BibitemOpen
  \bibfield  {author} {\bibinfo {author} {\bibfnamefont {K.}~\bibnamefont
  {Bagi}},\ }\bibfield  {title} {\bibinfo {title} {Stress and strain in
  granular assemblies},\ }\href {https://doi.org/10.1016/0167-6636(95)00044-5}
  {\bibfield  {journal} {\bibinfo  {journal} {Mechanics of Materials}\ }\textbf
  {\bibinfo {volume} {22}},\ \bibinfo {pages} {165} (\bibinfo {year}
  {1996})}\BibitemShut {NoStop}%
\bibitem [{\citenamefont {Bagi}(2006)}]{bagi_analysis_2006}%
  \BibitemOpen
  \bibfield  {author} {\bibinfo {author} {\bibfnamefont {K.}~\bibnamefont
  {Bagi}},\ }\bibfield  {title} {\bibinfo {title} {Analysis of microstructural
  strain tensors for granular assemblies},\ }\href
  {https://doi.org/10.1016/j.ijsolstr.2005.07.016} {\bibfield  {journal}
  {\bibinfo  {journal} {International Journal of Solids and Structures}\
  }\textbf {\bibinfo {volume} {43}},\ \bibinfo {pages} {3166} (\bibinfo {year}
  {2006})}\BibitemShut {NoStop}%
\bibitem [{\citenamefont {Schall}\ \emph {et~al.}(2007)\citenamefont {Schall},
  \citenamefont {Weitz},\ and\ \citenamefont
  {Spaepen}}]{schall_structural_2007}%
  \BibitemOpen
  \bibfield  {author} {\bibinfo {author} {\bibfnamefont {P.}~\bibnamefont
  {Schall}}, \bibinfo {author} {\bibfnamefont {D.~A.}\ \bibnamefont {Weitz}},\
  and\ \bibinfo {author} {\bibfnamefont {F.}~\bibnamefont {Spaepen}},\
  }\bibfield  {title} {\bibinfo {title} {Structural {Rearrangements} {That}
  {Govern} {Flow} in {Colloidal} {Glasses}},\ }\href
  {https://doi.org/10.1126/science.1149308} {\bibfield  {journal} {\bibinfo
  {journal} {Science}\ }\textbf {\bibinfo {volume} {318}},\ \bibinfo {pages}
  {1895} (\bibinfo {year} {2007})}\BibitemShut {NoStop}%
\bibitem [{\citenamefont {Heggen}\ \emph {et~al.}(2004)\citenamefont {Heggen},
  \citenamefont {Spaepen},\ and\ \citenamefont
  {Feuerbacher}}]{heggen_creation_2004}%
  \BibitemOpen
  \bibfield  {author} {\bibinfo {author} {\bibfnamefont {M.}~\bibnamefont
  {Heggen}}, \bibinfo {author} {\bibfnamefont {F.}~\bibnamefont {Spaepen}},\
  and\ \bibinfo {author} {\bibfnamefont {M.}~\bibnamefont {Feuerbacher}},\
  }\bibfield  {title} {\bibinfo {title} {Creation and annihilation of free
  volume during homogeneous flow of a metallic glass},\ }\href
  {https://doi.org/10.1063/1.1827344} {\bibfield  {journal} {\bibinfo
  {journal} {Journal of Applied Physics}\ }\textbf {\bibinfo {volume} {97}},\
  \bibinfo {pages} {033506} (\bibinfo {year} {2004})}\BibitemShut {NoStop}%
\bibitem [{\citenamefont {Callaghan}(1993)}]{callaghan_principles_1993}%
  \BibitemOpen
  \bibfield  {author} {\bibinfo {author} {\bibfnamefont {P.~T.}\ \bibnamefont
  {Callaghan}},\ }\href@noop {} {\emph {\bibinfo {title} {Principles of
  {Nuclear} {Magnetic} {Resonance} {Microscopy}}}}\ (\bibinfo  {publisher}
  {Oxford University Press},\ \bibinfo {year} {1993})\BibitemShut {NoStop}%
\bibitem [{\citenamefont {Torrey}(1956)}]{torrey_bloch_1956}%
  \BibitemOpen
  \bibfield  {author} {\bibinfo {author} {\bibfnamefont {H.~C.}\ \bibnamefont
  {Torrey}},\ }\bibfield  {title} {\bibinfo {title} {Bloch {Equations} with
  {Diffusion} {Terms}},\ }\href {https://doi.org/10.1103/PhysRev.104.563}
  {\bibfield  {journal} {\bibinfo  {journal} {Physical Review}\ }\textbf
  {\bibinfo {volume} {104}},\ \bibinfo {pages} {563} (\bibinfo {year}
  {1956})}\BibitemShut {NoStop}%
\end{thebibliography}
%

\end{document}


\pagenumbering{arabic}
\setcounter{page}{1}

\section*{\LARGE Lindemann unjamming of emulsions}

\vspace{1mm}
\section*{\Large Supplementary materials}
\vspace{2mm}

\section{Preparation and characterization of emulsions}

\subsection{Synthesis of nearly-monodisperse emulsions for DWS ---}\label{subsec:monodisp}

We prepare nearly-monodisperse emulsions of oil in water by swelling monodisperse particles of linear polystyrene made by a multi-step, seeded dispersion polymerization process~\cite{lee_influence_2008, jiang_seeding_2008, paine_dispersion_1990} with anisole.
We begin the process by synthesizing micrometer-size seeds using dispersion polymerization under bottle polymerzation conditions.
To do this we mix of 10 vol.\,\% styrene in ethanol, with 1.6 grams of poly(\textit{N}-vinylpyrrolidone) (Kollidon 30, BASF) stabilizer per hundred milliliters of liquid and 1\,wt.\,\% azobisisobutyronitrile (AIBN) to styrene in a round-bottom flask, then degas, evacuate, and seal it.
We then submerge the flask in an oil bath held at 70$^{\circ}$C and spin it at a constant rate of a few tens of RPM to provide gentle mixing.
We allow the reaction to proceed overnight, and wash these seed particles by repeatedly centrifuging and resuspending them in pure ethanol;
the growth steps proceed by adding these seeds to a solution similar to the one described above but with up to 40\,vol.\,\% styrene, and 0.5\,wt.\,\% azobis(cyanocyclohexane) (ACHN) and octanethiol to styrene.
The thiol reduces the average molecular weight of the new polymers, which reduces the viscosity of the final oil droplet and increases the swelling capacity of the particle, but interferes with the nucleation process.
By seeding the reaction with monodisperse particles much smaller than the final target size and using the more slowly decomposing ACHN radical initiator we avoid nucleating new particles and produce larger, monodisperse spheres composed of a lower molecular weight polymer.
We improve the uniformity of the resulting particles by splitting the total growth into three separate steps, thus preparing 4.2\,$\mu$m diameter particles.
After washing these particles with ethanol several times we resuspend them in a 2\,mM solution of SDBS and expose them to anisole.
The low molecular weight polymer chains provide enough osmotic pressure to swell the particles many times their starting volume, but we choose to swell them to a final diameter of 7.2\,$\mu$m.

\subsection{Characterization of nearly-monodisperse emulsions for DWS ---}
The \textbf{sedimentation velocity} of dilute droplets at this polymer/solvent composition is approximately 2\,mm/hr at room temperature, from which we estimate a \textbf{density difference} of $\delta\rho\!=$\,20\,kg/m$^3$ between the oil and water~\cite{mills_settling_1994}.
To find values of $\delta\rho$ that match the experimental conditions of the data shown in Fig.~2 of the main text we use the thermal expansion coefficients of anisole and water~\cite{haynes_compressibility_2013} and find values of $\delta\rho(T)$ at 27$^{\circ}$C, 31.5$^{\circ}$C, and 34.9$^{\circ}$C equal to 14\,kg/m$^3$, 11\,kg/m$^3$, and 8\,kg/m$^3$, respectively.

These values are essential to our computation of the gravitational stress, $\Pi(d)$, which we use to convert the measurements of $G(d)$ shown in Fig.~2B of the main text, into the values of $G(\Pi)$ shown in panel C of the same figure.
To do this we start from the mechanical equilibrium condition for the sediment:
\begin{equation}
\label{eq:Piofd}
\Pi(d)\! =\! g\,\delta\rho\,\int_{0}^d\phi(z)\,\mathrm{d}z
\end{equation}
where $g$ is the gravitational acceleration and $\delta\rho$ is the density difference between the two liquids, which is a measure of the depth dependence of the gravitational stress.
Assuming that $\phi(d)$ is nearly constant, $\phi(d)\!\approx\!\phi_c$, we reduce eq.~\ref{eq:Piofd} to $\Pi(d)=\delta\rho(T)\,g\,\phi_c\, d$.
We cannot measure $\phi_c$ directly in this sample, though the NMR densitometry measurements described in Section~\ref{sec:nmr} are designed to probe this very precisely, but taking $\phi_c$ to be the random close packing volume fraction of monodisperse spheres, $\phi_{RCP}\!\approx$\,0.635, provides an adequate approximation.
The particular value of $\phi_c$ we choose for this calculation will have no effect on how well the data shown in Fig.~2C of the main text collapse: at most slightly rescaling the bottom axis.

We use a ring tensiometer (Sigma 700, KSV) \cite{huh_rigorous_1975, huh_rigorous_1977} to measure the \textbf{interfacial tension} between bulk anisole-polystyrene mixtures and aqueous SDBS solutions at temperatures up to 40$^\circ$C, finding no appreciable variation in interfacial tension with temperature, and estimate that $\sigma\approx$\,5\,mN/m for the droplets in the emulsion.

\subsection{Sedimentation and equilibration of emulsion for DWS ---}

For the assumptions underlying DWS to be obeyed in a backscattering geometry the sample must be thick and wide enough to minimize the leakage of light through the back and sides of the container.
The size of the illuminated area must also many times larger than any microscopic correlation lengths so that the fluctuations of interference speckles are statistically independent, and thus capable of yielding true ensemble averages.
Consequently we hold the emulsion in 1\,cm thick by 3\,cm wide tubes, and expand the laser beam to a 1\,cm diameter.

For our quiescent sample we build a 20\,cm tall sediment of liquid droplets by allowing a 10\,vol.\,\% dispersion of droplets to settle inside a rectangular glass tube.
The Peclet number of these droplets is greater than 50, and crystallization of the droplets is not expected even at much lower sedimentation fluxes.
We accelerate the sedimentation rate further by tilting the column at a slight angle~\cite{boycott_sedimentation_1920}.
As the droplets settle we replace the clear supernatant with more droplet dispersion twice per day until the sediment reaches its final height, and seal the top of the tube with a glass plate and epoxy adhesive (Loctite E-120HP).
After the sample is sealed, we allow the sediment to slowly consolidate at room temperature for six months.

We transfer the emulsion column into the thermostatted box and mount it on a plate fixed to a vertical translation stage several months before performing any measurements.
We control the temperature inside the box by forcing air over a Peltier device that is controlled by an electronic thermostat (TED200, Thorlabs).
Another feedback controlled Peltier device thermostats the bottom plate of the illuminating laser (Compass 315M-50, Coherent Inc.).
The back plate of each Peltier device is attached to heat exchange blocks that are constantly flushed by a temperature controlled water recirculator.
The resonant cavity of the laser is cemented to the bottom plate of the device, and careful thermostatting of this element provides sufficient stabilization of the carrier frequency for the duration of each measurement.
We use thermistors to record the temperature at the bottom plate of the laser and of the air temperature inside the box, and find that the typical variation of the temperature of the bottom plate of the laser during a 20 hour measurement is below our measurement noise floor of $\pm$\,0.001$^{\circ}$C, and that the typical variation of the temperature inside the sample chamber is less than $\pm$\,0.01$^{\circ}$C.

\subsection{Preparation of bidisperse emulsion for mechanical rheology ---}\label{subsec:bidisp}

To measure the yield strain of a reference emulsion and fine-tune the accuracy of our optical microrheology apparatus we prepare an emulsion that can be measured using a conventional rheometer as well as optically.
We cannot carry out these measurements on the sealed, sedimented sample, and nearly-monodisperse emulsions are susceptible to shear crystallization, so we prepare a separate calibration sample with a shear modulus of $\approx$10\,Pa, composed of a bidisperse mixture of droplets.

To do this, we follow a similar protocol as described in Section~\ref{subsec:monodisp}.
We first synthesize 3.6\,$\mu$m and 4.5\,$\mu$m polystyrene particles using seeded dispersion polymerization, clean them, and suspend them in a 3\,mM solution of SDBS in water.
We then suspend approximately equal numbers of these particles in a 20\,vol.\,\% solution of D$_2$O with 3\,mM SDBS, and add enough anisole to swell the particles to five times their original volume.
After the anisole is absorbed by the particles, the emulsion is centrifuged at 15\,$g$ for several days while the clear subnatant is removed and stored.
We measure the shear modulus of the concentrated emulsion using oscillatory shear in a double-gap Couette geometry on a stress controlled rheometer (MCR501, Anton Paar), and replace enough of the stored subnatant to adjust the modulus to $\approx$\,10\,Pa.
Adding D$_2$O to the fluid changes the buoyant density of the swollen droplets in this suspension to $\approx$\,-\,4\,kg/m$^3$; therefore, the pressure difference between the top and bottom of the sample is $\lesssim$\,2\,Pa.

\subsection{Preparation of PDMS-TCE emulsion for NMR ---}
The resolution of our gradient echo sequence is best when measuring chemically identical nuclei; we thus use 20\,cSt PDMS as the source of proton signal, and D$_2$O as the source of deuterium.
The natural abundance of deuterium is low enough that we need not worry about deuterium signal arising from the PDMS, and the hydrogen content in high purity D$_2$O is also negligible.
The final concentration of SDBS is also small enough for its proton NMR signal to be negligible.

We begin by emulsifying PDMS using a cylindrical, porous-glass membrane (Shirasu Porous Glass, 1\,cm outer diameter, 2\,cm long, 3\,$\mu$m pore size).
We clean the membrane by soaking it in a potassium hydroxide solution and rinsing it with deionized water.
We then mount it on a stainless steel holder equipped with two o-rings and a fluid inlet.
To avoid trapping air in the membrane we flow CO$_2$ through the pores before and as we submerge the assembly into a 10\,mM  solution of sodium dodecyl sulfate (SDS) in water.
After the chamber is purged, we shut off the flow of gas and use a vacuum to draw the SDS solution into the membrane assembly.
We then connect the assembly to a syringe filled with the PDMS oil using a three way valve, and use a syringe pump to push oil through the membrane at a constant rate \cite{cheng_hydrophobic_2008, vladisavljevic_influence_2003}.  
When we operate the device at 1\,ml/hr we obtain droplets with a mean diameter of 10\,$\mu$m and a standard deviation of 0.8\,$\mu$m.

We increase the density of the droplets by swelling them with an equal volume of tetrachloroethylene, dividing the addition into three steps.
We first add one third the necessary TCE to the suspension of droplets and tumble them until the TCE has been absorbed.
These droplets are denser than D$_2$O, but not heavy enough to coalesce while still dispersed in water.  We allow the droplets to settle and replace the supernatant with a 3\,mM solution of SDBS in D$_2$O and add half of the remaining TCE.
We again tumble the emulsion until the TCE is absorbed, and repeat the process for the last aliquot of TCE.
To remove the residual SDS and H$_2$O from the initial emulsification we wash the emulsion several times: waiting for the droplets to settle, replacing the supernatant with a 2\,mM solution of SDBS in TCE saturated D$_2$O, and gently resuspending them.
Though this process removes most of the SDS, dodecanol impurities found in the SDS solution cannot be removed and reduce the interfacial tension of the droplets.

\section{Design, calibration, and characterization of light scattering apparatus}
To determine the low frequency mobility of the droplets we need to measure the long-time-plateau root mean squared displacement (RMSD) of the droplets.
We can estimate the order of magnitude of this RMSD using the equipartition of thermal energy relation for a bead embedded in an elastic solid:
\begin{equation*}
	\langle \Delta r^2\rangle\sim\frac{k_B T}{\pi R\,G}
\end{equation*}
where $G$ is shear modulus of the solid, and $R$ is the radius of the embedded particle.
For droplets several micrometers in diameter embedded in an elastic material with a shear modulus of a few Pa, we expect RMSDs of a few tens of nanometers.
This displacement is too small to accurately resolve using optical microscopy or single light scattering, but is also too large to resolve with transmitted light DWS.
Instead, we perform our DWS measurements using diffusively backscattered light, with which we can resolve RMSDs of up to 50\,nm.

\subsection{Design of DWS apparatus ---}

\begin{figure}
\centerline{\includegraphics[width=16.5cm]{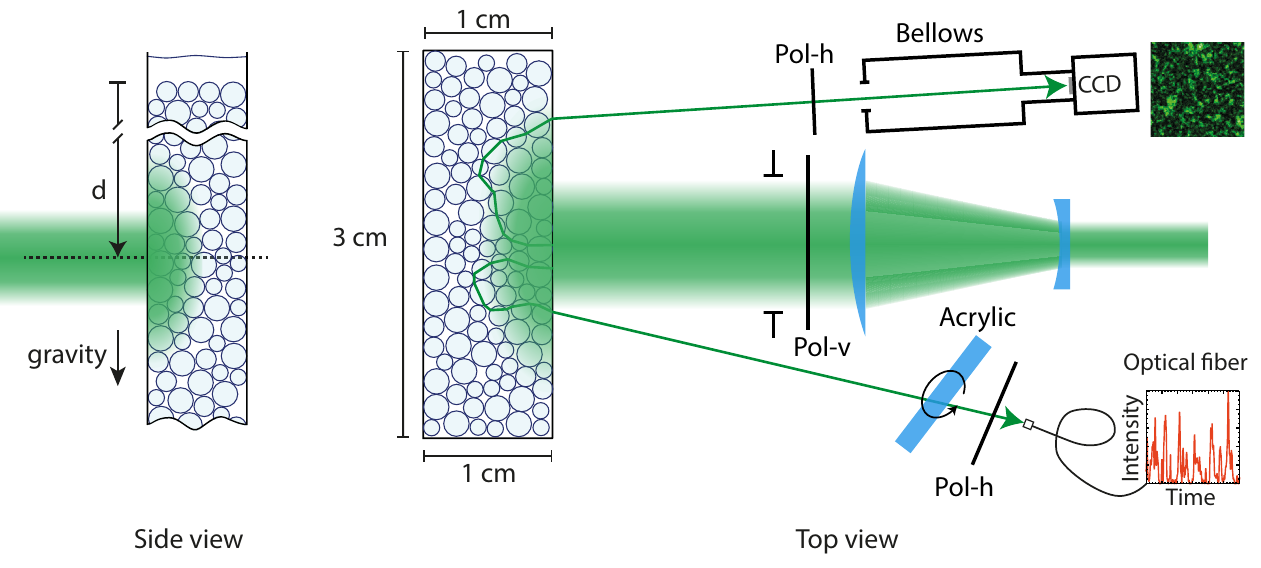}}
\caption{\textbf{Schematic representation of DWS apparatus} The emulsion is illuminated from one side with a 1\,cm diameter, vertically-polarized laser beam.
Horizontally polarized light diffusively reflected by the sample is collected by a single mode fiber and a CCD detector.
An acrylic disk placed in front of the fiber is rotated periodically at an oblique angle to refract the incident light, making it possible to sample many independent speckles.}
\label{fig:DWSschem}
\end{figure}

The true photon mean free path, $l$, of our emulsion should be comparable to the droplet size: as photons will scatter or be otherwise deflected every time they encounter an interface.
However, because our droplets are much larger than the wavelength of light, these photons will not be scattered isotropically but with a strong bias along the direction of incidence.
This then leads to a much longer ``effective" transport mean free path, $l^*$, that determines the distribution of photon path lengths in the material.
By comparing the relative opacity of our emulsion to a similarly sized block of PTFE---for which $l^*\!\approx$\,180\,$\mu$m for 532\,nm light \cite{rojas_photon_2011}---as well as the relative widths of the glowing halo around a laser focused on the front faces of both materials, we can tell the $l^*$ of our emulsion is comparable, though substantially smaller than that of our reference block of PTFE.
We thus estimate that $l^*$$\sim$\,100\,$\mu$m for our emulsion at this wavelength.

For the assumptions that underlie DWS microrheology to be valid, photons that arrive at a detector must have undergone long diffusive random walks within the sample, interacting with many independently fluctuating scatterers along the way, but diffusively reflected light includes paths as short as single reflections off the surface of the material.
However, because the typical path length over which the polarization of a photon is randomized, $l_p$, in this material is comparable to $l^*$, we exclude paths substantially shorter than $l^*$ when we illuminate the sample with vertically polarized light and collect horizontally-polarized backscattered light\cite{mackintosh_polarization_1989, xu_random_2005, bicout_depolarization_1994, rojas-ochoa_depolarization_2004}.
Moreover, since $l^*$ is over ten times larger than the diameter of a single droplet, even photons that follow paths as short as a single $l^*$ interact with many droplets before they exit the material.
A simplified schematic of our instrument is presented in Fig.~\ref{fig:DWSschem}.

For lag-times below a few tens of seconds, a hardware correlator (BI-9000AT, Brookhaven Instruments Corporation) cross-correlates photons detected by two avalanche photodiodes (APD, SPCM-AQRH, Perkin-Elmer) connected to a split, single-mode optical fiber.
The light detected by both APDs is collected from the same point, so the cross-correlated signal is statistically equivalent to an auto-correlated signal but without the effects of systematic detector and photon shot noise.
Temporal autocorrelations of the intensity for longer lag-times are computed from snapshots taken with a CCD camera (Retiga EXL, Qimaging, 63\,$\mu$s exposure time).
We eliminate stray light and reflections by placing an absorptive bellows in front of the CCD.

Scatterers embedded in an elastic medium have RMSDs that plateau to a finite value.  This localization restricts the variation in the phase acquired by the scattered photons and leads to speckle patterns that are not ergodic.
Proper ensemble averaging thus requires sampling many statistically independent speckles.
We describe our experimental solutions to this problem for camera and fiber-optic sensors in sections~\ref{subsubsec:camspec} and~\ref{subsubsec:fibspec}, respectively.

\subsubsection{Optimizing camera illumination and averaging ---}\label{subsubsec:camspec}
For large, area resolved detectors, such as a CCD, the solution to the non-ergodic sampling problem is straightforward: the sensor simultaneously records the intensity of thousands of independent speckles that can be cross-correlated and averaged.
The size of speckles arriving at a detector placed far from the illuminated sample is determined by the van Cittert-Zernike theorem:
\begin{equation*}
	d\sim\frac{\lambda\,z}{D}
\end{equation*}
were $d$ is the spatial correlation length of interference speckles of quasi-monochromatic light of wavelength $\lambda$, measured a distance $z$ away from an incoherent source of size $D$.
Though the spatial correlation length of light at the surface of the emulsion is larger than $\lambda$, neighboring speckles recorded by the camera will be statistically independent when the illuminated area is sufficiently large.
The correlation size and shape of these speckles can be approximated from spatial auto-correlations of recorded frames.
This approximation is best when the active area of each pixel is much smaller than the speckle correlation length, as individual pixel values are equivalent to noisy integrals over the pixel area rather than true measurements of local intensity.
However, the approximation is adequate when the speckle is substantially larger than a single pixel, as is the case for all our measurements, and the true spatial correlation should be recoverable by deconvolving the autoconvolution of the pixel active area from the measured correlation.

\begin{figure}
\centerline{\includegraphics[width=16.5cm]{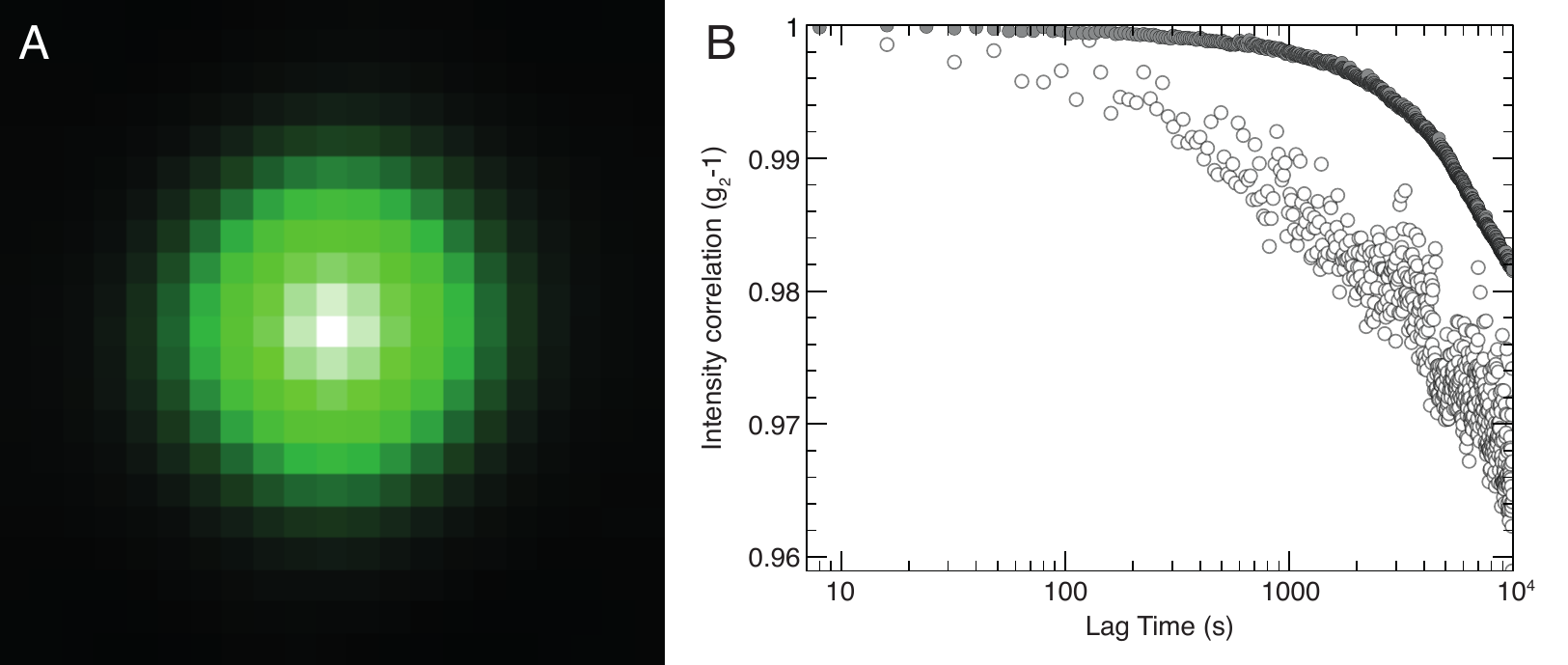}}
\caption{\textbf{DWS measurements of backscattered light from sintered glass beads and PTFE block}
(A) Central portion of the spatial cross-correlation of laser light scattered from a sintered pack of borosilicate glass beads computed from frames gathered 8\,s apart.
(B) Intensity correlations for light scattered from a sintered glass bead pack (gray circles) and a PTFE block (open circles).}
\label{fig:DWScalib}
\end{figure}

Spatial auto- and cross-correlations of uniformly illuminated frames should have a single peak at the origin standing on a perfectly flat background, as depicted in Fig.~\ref{fig:DWScalib}A: the with of this peak is the spatial correlation length.
We adjust the distance of the camera from the sample to make the correlated area of each speckle comparable to that of ten pixels: which provides $\sim\!10^5$ independent speckles per snapshot.
The pixel size is still comparable to the spatial correlation length of the speckle pattern, so the $\tau\!\to$\,0 limit of the intensity-intensity autocorrelation, $g_2(\tau)=\frac{\langle I(t)\,I(t+\tau)\rangle}{\langle I(t)\rangle^2}$, is not exactly 2, but lies between 1.8 and 1.9.
Frame auto-correlations will have an additional, delta-correlated peak centered at the origin that corresponds to the variance of the pixel and photon shot noise.

The position and shape of the correlation peak reflects the effects of mechanical drift and vibrations.
Peaks from auto-correlated frames gathered with a sufficiently short exposure time are not smeared or distorted by motion averaging, but such motions are reflected in cross-correlated frames as displacements of the peak centroid.
We fit the correlation peak to a paraboloid and compute the peak shape, centroid, and amplitude for all frame pairs: discarding pairwise correlations between frames that show displacements greater than one or two pixels.
More dramatic perturbations produce more severe distortions, so we discard pairwise correlations that show correlation peaks with ellipticity or radii of gyration several standard deviations greater than the mean.
This filtering process can be important for discriminating peaks in more mobile samples, where the peak correlation intensity may fall below the noise floor of our fitting algorithm.

\subsubsection{Optimizing fiber-optic illumination and averaging ---}\label{subsubsec:fibspec}
Though it is not possible to simultaneously average many speckles with a single-mode fiber-optic detector, we can provide unbiased sampling by manipulating the statistical properties of the light incident on the fiber.
This can be done by placing another multiply-scattering element between the sample and the detector, but this reduces the intensity and scrambles the polarization of the detected light.

Instead, we deflect the incoming light using a large, transparent acrylic disk placed at an oblique angle to the optical axis of the fiber, as depicted schematically Fig.~\ref{fig:DWSschem}.
The thickness of the disk and the angle of its normal with the optical axis of the fiber determine the absolute deflection distance of the incident light.
Spinning the tilted disk along the optical axis of the fiber causes the surface normal of the disk to precess, rigidly displacing the light arriving at the fiber tip and thus making it possible to sequentially sample hundreds of independent speckles.
As before, we place the fiber far enough away from the emulsion to sample only independent speckles, and rotate the disk by an angle large enough to randomize the intensity after every step.
A larger angle of incidence will deflect the light more, but will also affect the polarization of the transmitted light more.
We chose a disk 17\,mm thick, with a normal pointed 10 degrees away from the optical axis of the fiber, and rotate it by 2.7 degrees every 90 seconds.
The abrupt change between independent speckles introduces a windowing artifact that appears as an artificial decorrelation of the measured $g_2(\tau)$;
however, if adjacent measurement angles sample independent speckles, this effect can be removed by dividing the measured correlation by 1\,$-\tau/\mathrm{T}$, where $\mathrm{T}$ is the time between steps.

\subsection{Apparatus Stability and Calibration}
\subsubsection{Decorrelation of light scattered from static samples}
We use a sintered pack of borosilicate beads and a block of polytetrafluoroethylene (PTFE, Teflon) to test the stability of our instrument.
The speckle patterns of light scattered from these materials should be perfectly static; however, vibrations, temperature driven dilations of the sample, and drift of the laser frequency will lead to changes in the interference pattern, and thus to decorrelation.
The thermal expansion coefficient of the glass beads is negligible, but that of the PTFE block is comparable to that of our oil droplets.
Consequently, by measuring the intensity autocorrelation of light backscattered from these two materials it is possible to separate the effects of fluctuations in the temperature of the enclosure from the effects of vibrations and laser fluctuations.
The intensity correlations measured for both materials are almost perfectly static, but those measured for the PTFE block decay somewhat more than those for the sintered beads, as shown in Fig.~\ref{fig:DWScalib}B.

Thermal fluctuations in these calibration media should be perfectly reversible.
By contrast, fluctuations in the temperature of the emulsion are converted into uniaxial strains by the large difference in thermal expansion coefficients between the glass tube and the oil droplets.
These strains can lead to irreversible rearrangements of the droplets and thus to irreversible speckle intensity decorrelation;
however, since the temperature in enclosure does not vary by more than $\pm$\,0.01$^{\circ}$C during a typical measurement, the resulting shear strains are limited to amplitudes smaller than 10$^{-5}$ ($\gamma\!\approx\!\alpha\,\delta T$).
By comparison, the yield strain at the liquid-solid interface, estimated from the ratio of plateau RMSD to droplet diameter, is $\gamma_y\approx\frac{10\,\mathrm{nm}}{7.2\,\mu\mathrm{m}}\approx10^{-3}$.

\subsubsection{Calibration of $\beta$ and $\gamma$ parameters for DWS microrheology analysis}
The elastic moduli of the sintered bead pack and PTFE block and the size of their scattering structures are too large for the amplitude of their thermally induced vibrations to exceed picometer scales.
By contrast, the positions of micrometer scale droplets belonging to an emulsion with a shear modulus of a few Pa can be expected to fluctuate several nanometers.
For diffusively backscattered light, the time dependence of the speckle intensities is related to these displacements by the following set of relations:
\begin{subequations}
\begin{align}
	g_2(\tau)&=\frac{\langle I(t+\tau)I(t)\rangle}{\langle I(t)\rangle^2} \label{DWSrln_a}\\
	&=\beta\, g_1(\tau)^2+1 \label{DWSrln_b}\\
	\sqrt{\langle \Delta r(\tau)^2\rangle}&=-\frac{\log(g_1(\tau))}{\gamma\,k} \label{DWSrln_c}\\
	G&=\frac{k_B T}{\pi R\, \langle\Delta r(\tau\! \to\! \infty)^2\rangle}\label{DWSrln_d}
\end{align}
\end{subequations}
The angle brackets in (\ref{DWSrln_a}) refer to averages over the CCD detector pixels or over the sequence of speckles sampled by the optical fiber.
When the fluctuations in the electromagnetic field are gaussian, the Siegert relation (\ref{DWSrln_b}) can be used to recover the field-field correlation, $g_1$, from the measured intensity-intensity correlation, $g_2$,~\cite{berne_dynamic_2000}.
The parameter $\beta$ is equal to 1 when a single component of the the intensity is sampled at a single point in space.
The value of $\beta$ decreases when: the area of the detector is comparable to the speckle size; more than one polarization mode is detected; the incident light has more than one statistically independent mode.
The single mode fiber and polarizer effectively filter out all but a single statistically independent speckle, and $\beta$ for this measurement is almost exactly equal to 1.
However, the finite size of the CCD pixels lead to some spatial averaging that reduces the value of $\beta$ to $\approx$\,0.8\,--\,0.9.

When the transport of light inside the opaque material may be assumed to be diffusive and absorption may be neglected, (\ref{DWSrln_c}) provides a direct relation between the speckle fluctuations and the mobility of the individual scatterers.
Here, $k\!=\!\frac{2\pi\,n_{\mathrm{eff}}}{\lambda}$ is the effective wavenumber of the laser light inside of the material, and $\gamma$ is a parameter that measures the typical length of the photon paths in units of $l^*$, which typically lies between 1.5 and 2.5.
This approximation assumes that the scattered correlation is dominated by contributions from long paths; however, the phase accumulated in longer paths decorrelates more quickly than that from shorter ones: (\ref{DWSrln_c}) is thus only valid when $-\log(g_1\!)$ is not much greater than 1.  For $\lambda\!=$\,532\,nm and $\gamma\!\approx$\,2, this condition limits resolvable displacements to $\lesssim$\,50\,nm.

The value of this RMSD is connected to the shear modulus of the emulsion by the equipartition relation of elastic energy (\ref{DWSrln_d}), which is the long lag-time limiting form of the the Generalized Stokes-Einstein relation~\cite{mason_diffusing-wave-spectroscopy_1997}.
The frequency dependence of the viscoelastic shear modulus may be obtained from the analytic continuation of the Laplace transform of $\langle \Delta r(\tau)^2\rangle$ along the imaginary frequency axis, but this process requires fitting the data to a suitable analytical model over $\langle \Delta r(\tau)^2\rangle$ several decades in $\tau$.
Yet it is possible to extract the low frequency shear modulus, $G$, of an elastic material by considering only the $\tau\!\to\! \infty$ limit of $\langle \Delta r(\tau)^2\rangle$.
However, to compute this limiting value from (\ref{DWSrln_c}) we must first determine the value of $\gamma$, which we fix by comparing DWS measurements to mechanical rheology.

\begin{figure}
\centerline{\includegraphics[width=16.5cm]{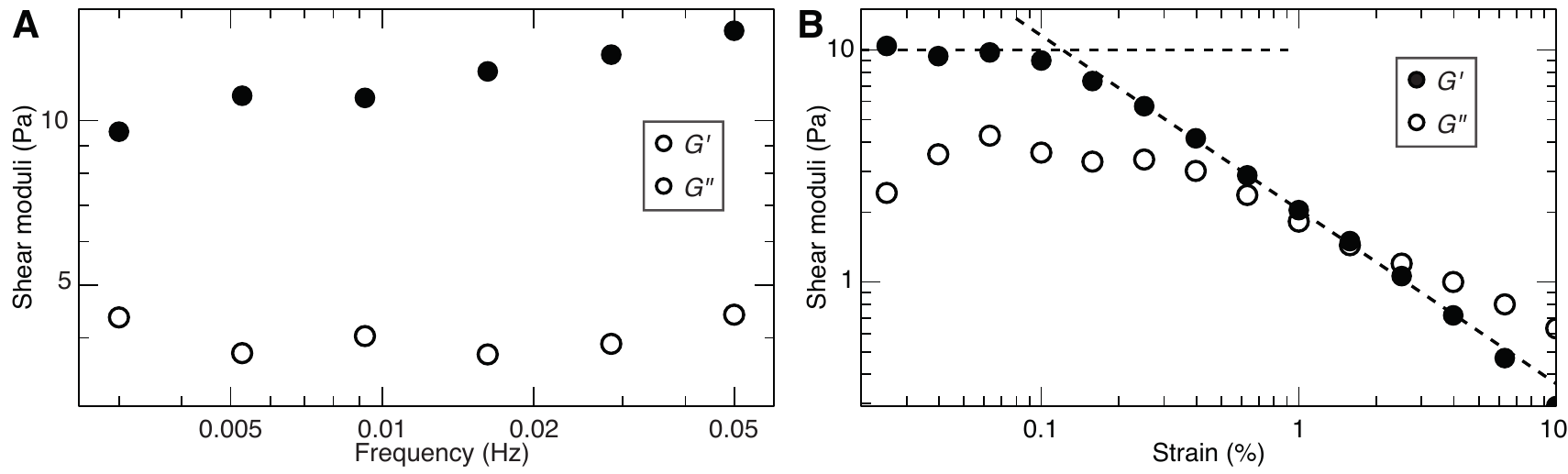}}
\caption{\textbf{Oscillatory rheology measurements of bidisperse calibration emulsion}
(A) Real and imaginary parts of the viscoelastic modulus of a compressed emulsion measured with oscillatory frequency sweeps with an oscillation amplitude of 0.05\,\%.
(B) Real and imaginary parts of the response to 0.005\,Hz frequency oscillatory strains with increasing amplitude.}
\label{fig:rheosweeps}
\end{figure}

We cannot carry out these measurements on the sealed, sedimented sample, so we prepare a separate calibration sample with a bidisperse mixture of droplets and a shear modulus of $\approx$10\,Pa, as described in Section~
\ref{subsec:bidisp}.
Swept frequency measurements performed by a stress controlled rheometer (MCR501, Anton Paar) show a nearly frequency independent storage modulus between 0.003\,Hz and 0.05\,Hz. Swept oscillatory strain measurements at 0.005\,Hz show linear, elastic behavior for strains below 0.1\,\% (Fig.~\ref{fig:rheosweeps}).

\begin{figure}
\centerline{\includegraphics[width=8.2cm]{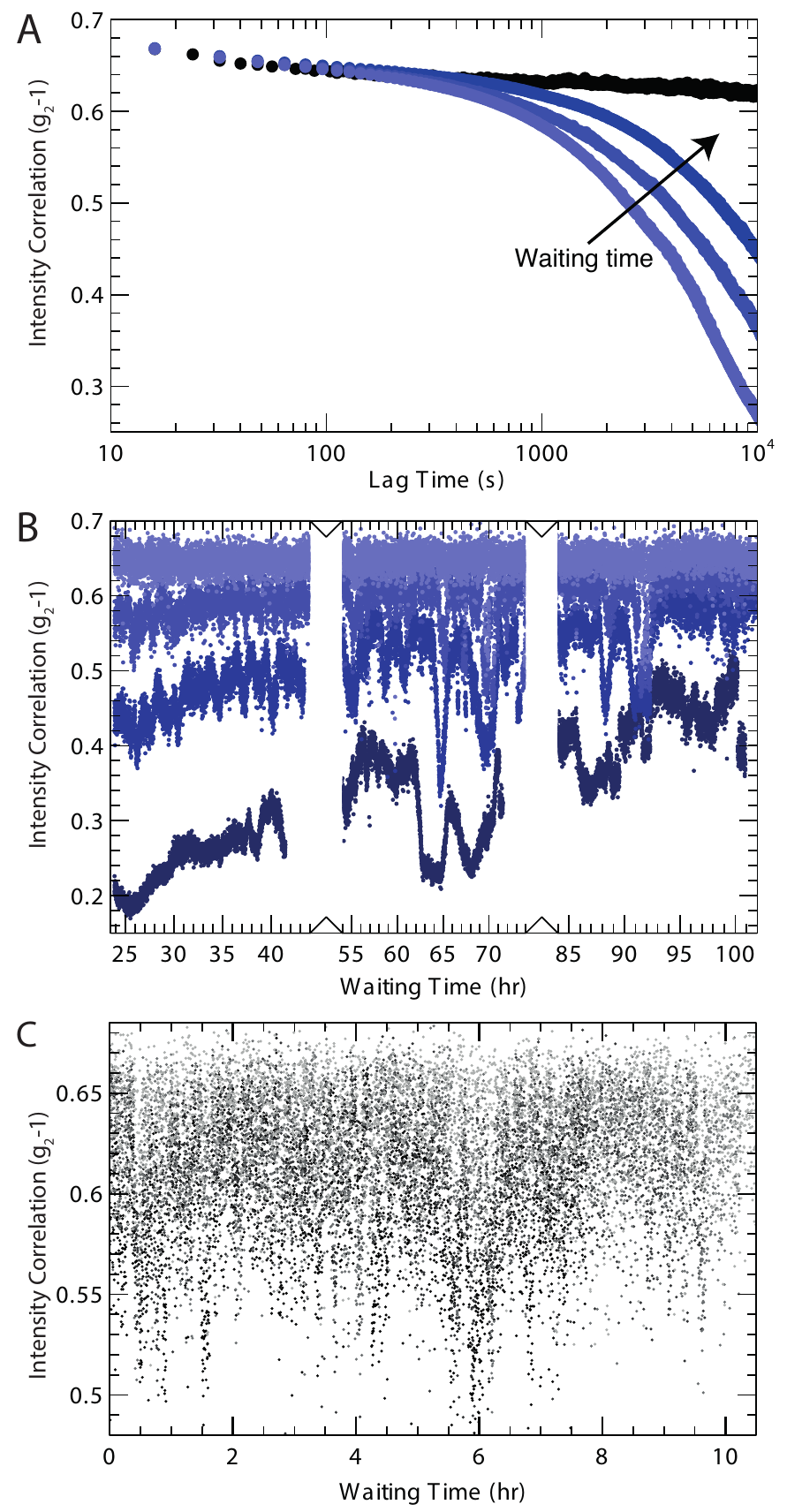}}
\caption{\textbf{Time resolved DWS correlations of aging and equilibrated emulsion samples}
(A) Binned intensity correlations for samples aged for one, two, and three days (light to dark blue), as well as a sample aged for several months.
(B) Time resolved correlations (TRC) for the aging sample measured at lag-times of 10$^2$\,s, 10$^3$\,s, 3\,$\cdot$\,10$^3$\,s, and 10$^4$\,s (light to dark blue).
(C) Time resolved correlations (TRC) for the equilibrated sample measured at lag-times of 10$^2$\,s, 10$^3$\,s, 3\,$\cdot$\,10$^3$\,s, and 10$^4$\,s (gray to black).}
\label{fig:aging}
\end{figure}

We take 15\,mL of this reference sample and seal it in a 4\,cm wide, 1\,cm deep, rectangular, glass cell.
We place this sample in our temperature controlled enclosure, adjust the ambient temperature to 22$^{\circ}$C, and wait one day before recording the scattered light intensity.  To track the effects of sample aging we perform 20\,hr long measurements on three consecutive days.
Binning the intensity correlation data from these three measurements leads to the three blue correlation functions shown in Fig.~\ref{fig:aging}A.
All three measurements agree for lag-times shorter than $\approx$100\,s, but the effects of aging are clearly visible in the data corresponding to longer lag-times.
Measurements performed on aging samples are not stationary, however; the average described in (\ref{DWSrln_a}) should thus not be done over the waiting time, $t$, but over independent speckles.
We compute a time resolved correlation (TRC) using the speckle images gathered with the CCD as~\cite{cipelletti_time-resolved_2003}:
\begin{equation}
\label{TRCeq}
	g_2(\tau,t)=\frac{\langle I(t+\tau,x)I(t,x)\rangle_x}{\langle I(t+\tau,x)\rangle_x\langle I(t,x)\rangle_x}
\end{equation}
Where $\langle\cdot\rangle_x$ denotes spatial frame averages.
We thus plot the waiting time dependence of the intensity correlation for lag-times of 10$^2$\,s, 10$^3$\,s, 3\,$\cdot$\,10$^3$\,s, and 10$^4$\,s in Fig.~\ref{fig:aging}B.
The correlation values for $\tau\!=$\,10$^2$\,s appear stationary, while values for $\tau\!=$\,10$^3$\,s show some time dependence: reaching a fairly steady plateau, with occasional decorrelation avalanches, after 2 days of aging.
Correlation values for $\tau\!=$\,3\,$\cdot$\,10$^3$\,s may have reached a plateau after 4 days, but show much larger decorrelation avalanches.
Correlation values for $\tau\!=$\,10$^4$\,s, however, show pronounced and persistent avalanches and may take many more days to reach a final plateau.
By contrast, the TRCs of emulsions that have aged for several months are almost perfectly stationary (Fig.~\ref{fig:aging}C).

Instead of waiting the weeks or months that would be necessary to equilibrate this reference sample, we find the measurement of $g_2(\tau)$ from the equilibrated sediment that best matches the reference data up to $\tau\!\approx$\,10$^3$\,s, and proceed with the analysis.
If we replace the $\tau\!\to\!\infty$ limit in (\ref{DWSrln_d}) with $\tau\!=$\,10$^4$, and use (\ref{DWSrln_a}-c) to compute $\langle \Delta r(\tau\!=$\,10$^4)^2\rangle$, we can equate the the shear modulus of the reference sample with that of the aged sample by choosing $\gamma\!=$\,2.27.
This value of $\gamma$ is within the commonly observed range for this parameter; but, given the uncertainty of the measurement, we choose to fix $\gamma\!\equiv$\,2 to compute the shear moduli of the quiescent emulsion.
The difference between shear moduli computed with either value of $\gamma$ is less than 30\%, and simply rescales $G(\Pi)$ by a constant factor.

\subsubsection{Scattering from fluctuating droplet interfaces}
Thermal fluctuations cause not only fluctuations of the droplet center positions, but also the fluctuation of the droplet interfaces.
Typical fluctuations in droplet surface area caused by these shape distortions are O($\frac{k_B\,T}{\sigma}\!\approx$\,1\,nm$^2$), where $\sigma$ is the interfacial tension of the droplets.
To illustrate how this change in area translates into a distortion of the sphere we consider the lowest order deformation mode: the volume preserving transformation from a sphere of radius $R$ to an ellipsoid with $R_z$\,=$R$\,+\,$\epsilon$ and $R_x$\,=\,$R_y$\,=\,$R\sqrt{\frac{R}{R+\epsilon}}$.
To leading order, this transformation increases the area of the droplet by $\delta A\!=\!8\pi\epsilon^2/5$.
Consequently, the average magnitude of this distortion is $\langle\epsilon^2\rangle^{1/2}\approx\sqrt{\frac{5k_B\,T}{8\pi\,\sigma}}\approx$\,0.4\,nm.

Nevertheless, since positional fluctuations are not strongly correlated to the interfacial fluctuations, these two fluctuations contribute additively to $\langle \Delta r(\tau)^2\rangle=\langle \Delta r(\tau)^2\rangle_{pos}+\langle \Delta r(\tau)^2\rangle_{int}$, with $\frac{\langle \Delta r(\tau)^2\rangle_{int}}{\langle \Delta r(\tau)^2\rangle_{pos}}\!\ll$\,1.
Moreover, the lifetime of these interfacial fluctuations, $\tau\sim\frac{R\,\eta}{\sigma}\approx$\,5\,$\mu$s, is too short to be resolved by the 63\,$\mu$s exposure time of our images~\cite{gang_shape_1994}.
The measured pixel intensities thus average over many configurations of the interfacial fluctuations, and these displacements are absorbed into an effective $\beta$ that is reduced, from 0.8\,--\,0.9 in the case of sintered beads, to $\approx$\,0.7\,--\,0.8 for our emulsions.
Dividing the measured intensity correlations by this effective $\beta$ subtracts and thus eliminates the contribution of interfacial fluctuations from our measurement of $\langle \Delta r(\tau)^2\rangle$.

\section{Long lag-time $g_2(\tau)$ data}

\begin{figure}
\centerline{\includegraphics[width=8.8cm]{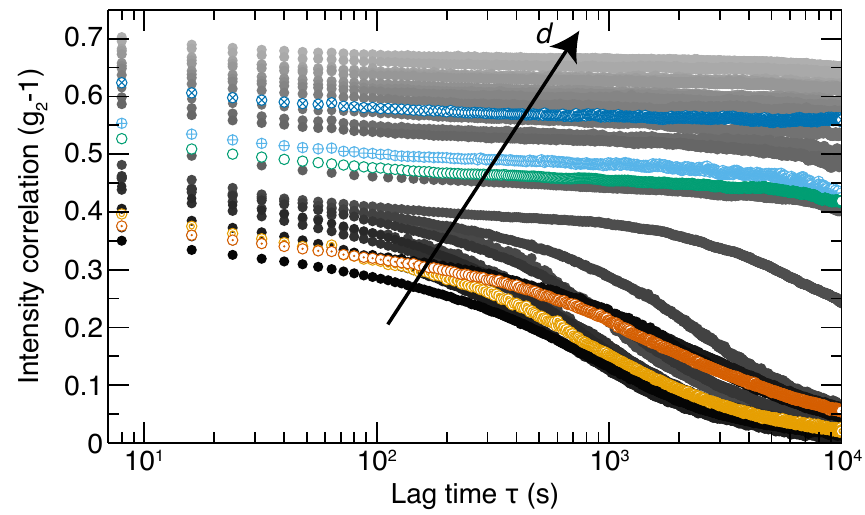}}
\caption{\textbf{Long lag-time $g_2(\tau)$ data for sediment at 31.5$^\circ$C}
Intensity correlations computed from snapshots recorded by the camera at different vertical depths into the sediment.
Symbol, color, and shading of each correlation shown here corresponds to the equivalently labeled $G(d)$ and $G(\Pi)$ points shown in Fig.~2 of the main text.
}
\label{fig:fullg2}
\end{figure}

Fig.~1B of the main text shows five correlation functions spanning the full range of available lag times that illustrate the jammed and entropic behaviors that characterize our data.
However, we recorded a total of 29 such correlations from the sediment equilibrated at 31.5$^\circ$C, which we then used to compute the data shown in Fig.~2 of the main text.
For the sake of completeness, we present all the correlations computed from snapshots recorded by the camera at this temperature in Fig.~\ref{fig:fullg2}.

With the exception of a single measurement, these correlations clearly sort themselves into rigid solid --- $g_2(\tau\to\infty)$ reaches a plateau --- and entropic --- $g_2(\tau\to\infty)$ is indistinguishable from 0 --- behaviors.
The behavior of the outlier is likely due to the mixing of light scattered from jammed and entropic regions, as the 1\,cm diameter beam illuminates a region where the mechanical properties of the sample vary rapidly.

\section{Connection between microscopic strains and particle MSDs}

\begin{figure}
\centerline{\includegraphics[width=12cm]{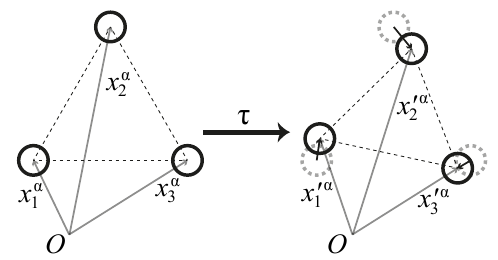}}
\caption{\textbf{Microscopic strain of a fluctuating simplex}
The positions of D+1 particle centers, $x_i^\alpha$, measured from an arbitrary origin, $O$, define a D-dimensional simplex (dashed black lines).
After a time $\tau$, the Brownian motion will displace the vertex particles to new positions, $x$'$_i^\alpha$, translating and distorting the simplex defined by them.
}
\label{fig:shearmsd}
\end{figure}

The coordinates of all droplet centers, $x_i^\alpha$, can be used to subdivide the entire volume of an emulsion into non-overlapping tetrahedra using algorithms like the Delaunay tessellation.
The position and shape of these tetrahedra will fluctuate along with the random Brownian motion of droplet positions (Fig.~\ref{fig:shearmsd}).
Large enough particle displacements can cause tetrahedra defined from an initial coordinate snapshot to overlap or otherwise fail to be a proper subdivision of space, but the distortion of these tetrahedra can be an excellent measure of material strain \cite{bagi_stress_1996,bagi_analysis_2006}.

For individual tetrahedra (or D-dimensional simplexes), it is always possible to find an affine transformation that maps the initial set of coordinates, $x_i^\alpha$, to those measured at some later time, $x_i^{\prime\alpha}$:
\begin{equation*}
	x_i^{\prime\alpha}=\sum_\beta{}F^\alpha_\beta{}x_i^\beta+c^\alpha
\end{equation*}
Were $F^\alpha_\beta$ is the deformation gradient tensor and $c^\alpha$ is a rigid translation. Defining $\delta x_i^{\alpha}\equiv{}x_i^{\prime\alpha}-x_i^{\alpha}$, makes it possible to rewrite this expression as D$\cdot$(D+1) linear equations for the unknown elements of $F-I$ and $c^\alpha$:
\begin{equation*}
	\delta{}x_i^{\alpha}=\sum_\beta{}(F^\alpha_\beta-\delta^\alpha_\beta)x_i^\beta+c^{\alpha}
\end{equation*}
Which clearly shows that $F-I$ and $c^\alpha$ are directly proportional to the magnitude of droplet displacements.
Choosing the origin, $O$, such that $\sum_ix_i^\alpha=0$ simplifies the equations further, guaranteeing that $c^\alpha=\frac{1}{4}\sum_i\delta{}x_i^\alpha$.
However, regardless of the choice of origin, $F$ can always be decomposed into the product of an orthogonal rotation matrix and a symmetric stretch tensor, while its symmetric product defines a Cauchy-Green deformation, $C=F^T\!\cdot\!F$, and the Green-Lagragian strain, $E=\frac{1}{2}(C-I)$, tensors.

For externally applied strains, the values of $F-I$ measured at every tetrahedron may not be precisely identical, but they will be coherent over macroscopic length-scales.
By contrast, because $\langle \delta{}x_i^{\alpha}\rangle=0$ for quiescent materials, $\langle F-I\rangle$ will also be zero and the values of $F-I$ for individual tetrahedra will be random variables displaying local spatial correlations \cite{hinch_application_1975,Eshelby1957,Eshelby1959}.

The random, largely incoherent nature of these local strain fluctuations helps explain why strong materials that require large, spatially coherent stresses to fail are stable at room temperature.
However, there is ample evidence that suggests that the yielding of emulsions and other jammed solids is mediated by localized plastic rearrangements, which may be cooperative, rather than by a coherent, macroscopic process like fracture.
This is similar to the concept of Shear Transformation Zones in glassy materials, and our finding from eq.\,4 of the main text that the activation volume of our transition appears to be only $V_0\!\approx6\!\cdot\!\frac{4\pi}{3}R^3$, is consistent with this interpretation.

To demonstrate the connection between MSD and strain more concretely we consider the RMS local strain of 4 particles arranged in a regular tetrahedron with side length $2R$, subjected to uncorrelated, zero-mean Gaussian noise with variance $\sigma^2$.
In a coordinate system where $x^\alpha_{1,2}=(\pm R,0,-R/\sqrt{2})$ and $x^\alpha_{3,4}=(0,\pm R,R/\sqrt{2})$, we obtain:

\[F-I = \frac{1}{2R}\left( \begin{array}{ccc}
\delta{}x_1^x-\delta{}x_2^x & \delta{}x_3^x-\delta{}x_4^x & \frac{-\delta{}x_1^x-\delta{}x_2^x+\delta{}x_3^x+\delta{}x_4^x}{\sqrt{2}}  \\
\delta{}x_1^y-\delta{}x_2^y & \delta{}x_3^y-\delta{}x_4^y & \frac{-\delta{}x_1^y-\delta{}x_2^y+\delta{}x_3^y+\delta{}x_4^y}{\sqrt{2}} \\
\delta{}x_1^z-\delta{}x_2^z & \delta{}x_3^z-\delta{}x_4^z & \frac{-\delta{}x_1^z-\delta{}x_2^z+\delta{}x_3^z+\delta{}x_4^z}{\sqrt{2}} \end{array} \right) .\]

The expression for the Green-Lagrange strain, $E$, follows trivially from this but is too cumbersome to reproduce here;
however, because it is a quadratic function of $F$, it does have a non-zero ensemble average, $\langle E^\alpha_\beta\rangle=\frac{3\sigma^2}{4R^2}\delta^\alpha_\beta$, which is an isotropic dilation and thus not involved in yielding \textit{per se}.
However, we can rewrite this tensor using Voigt notation $\tilde{E}_\alpha=(E_{xx},E_{yy},E_{zz},2E_{yz},2E_{xz},2E_{xy})$, and compute its covariance:

\[\langle(\tilde{E}-\langle\tilde{E}\rangle)\otimes(\tilde{E}-\langle\tilde{E}\rangle)\rangle = \frac{\sigma^2}{R^2}\left( \begin{array}{cccccc}
\frac{1}{2}+\frac{3\,\sigma^2}{8 R^2} & 0 & 0 & 0 & 0 & 0 \\
0 & \frac{1}{2}+\frac{3\,\sigma^2}{8 R^2} & 0 & 0 & 0 & 0 \\
0 & 0 & \frac{1}{2}+\frac{3\,\sigma^2}{8 R^2} & 0 & 0 & 0 \\
0 & 0 & 0 & 1+\frac{3\,\sigma^2}{4 R^2} & 0 & 0 \\
0 & 0 & 0 & 0 & 1+\frac{3\,\sigma^2}{4 R^2} & 0 \\
0 & 0 & 0 & 0 & 0 & 1+\frac{3\,\sigma^2}{4 R^2}  \end{array} \right) .\]
And the rotationally invariant combination $\sqrt{\langle E_{yz}^2+E_{xz}^2+E_{xy}^2\rangle}\approx\frac{\sqrt{3}\,\sigma}{2R}$ is of the same order of magnitude as the quantity we postulated sets the maximum RMSD in our modified Lindemann criterion $\sqrt{\langle \Delta r^2(\Pi^*)\rangle}/2R=\gamma_y(\Pi^*)$.

Though, to our knowledge, the spatial correlations between these tetrahedral strains have not been measured either experimentally or in simulation, the deformations of one tetrahedron will necessarily influence the deformations of its neighbors.
Consequently, while we cannot make a fully quantitative prediction without such microscopic measurements, it is not hard to see how localized strain fluctuations could cause materials with exceedingly small yield strains to fail under thermal stresses alone.

\section{Design, calibration, and characterization of MRI densitometer}\label{sec:nmr}

\begin{figure}
\centerline{\includegraphics[width=8.8cm]{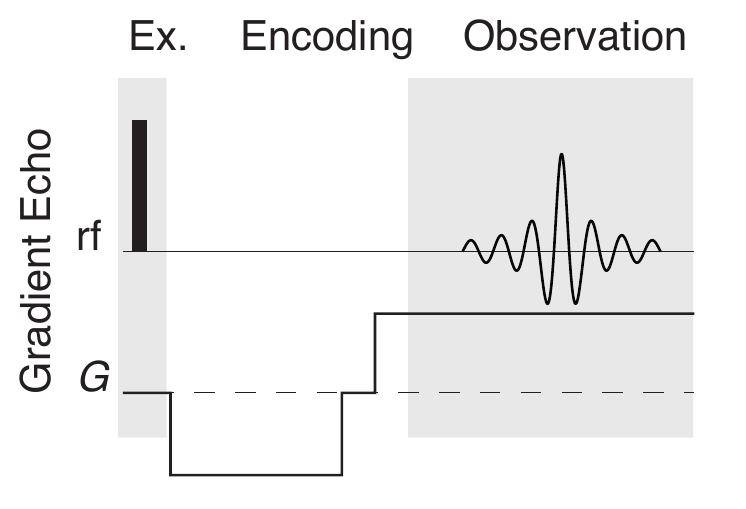}}
\caption{\textbf{Simplest gradient echo} This sequence uses a gradient to dephase the spins in a sample, then reverses it to produce a burst of magnetization.}
\label{fig:seqschem}
\end{figure}

Magnetic imaging or sectioning techniques rely on pulsed magnetic field gradients that temporarily shift the resonant frequency of the measured spins based on their position~\cite{callaghan_principles_1993}.
Though the assumption that these coils apply a perfectly uniform gradient along the direction of the principal field is unphysical, within the approximations of the Bloch-Torrey equations \cite{torrey_bloch_1956}, the phase of spins precessing in a constant magnetic field with an additional gradient, $\vec{B}(t)\!=\!B_0\hat{z}+ G(t)\,(-x\hat{x}/2-y\hat{y}/2+z\hat{z})$, will depend on their position in space:
\begin{equation*}
	\frac{\textrm{d}}{\textrm{d}t}\psi(\vec{r},t)=\omega_0+\gamma_0\,G(t)\,z
\end{equation*}
where $\gamma_0$ is the gyromagnetic ratio of the measured nucleus, and $\omega_0$ is the difference between the natural precession frequency of the spin and that of the chosen rotating reference frame.
Gradient echoes are the simplest of these measurement techniques, and they have the highest resolution and sensitivity, but they can only operate in one spatial dimension and work best when all the interrogated nuclei have exactly the same resonant frequency, $\gamma_0B_0$.
Here all the spins in the measurement volume are excited using a broadband pulse, the gradient is then turned on for some time, and subsequently reversed (Fig.~\ref{fig:seqschem}).
The first gradient pulse winds the phase of precessing spins belonging to different z-positions in the sample along the gradient axis, while the second gradient pulse unwinds them: causing a short burst of precessing magnetization once the phases of the spins are refocused.
If the amplitude and phase of the magnetization is recorded for the entire duration of the second gradient pulse, its Fourier transform corresponds to the density profile.

This technique assumes that each frequency component of the spectrum corresponds to a separate spatial positions; however, when two or more chemically distinct nuclei are present, the resulting spectrum is a convolution of the spatial distribution of spins with their chemical frequency spectrum: know as a chemical shift artifact~\cite{callaghan_principles_1993}.
A similar effect is seen when the magnetic coupling between atoms that share a bond, typically $^{13}$C and $^1$H, splits the signal of the observed nucleus into two or more peaks.
The $\approx$\,1.11\% natural abundance of $^{13}$C atoms makes this effect rather small for $^1$H spectroscopy, but is significant enough to require correction.
This is commonly done using broadband heteronuclear decoupling sequences~\cite{levitt_broadband_1982}, which use rf-pulses tuned to the resonant frequency of the unobserved nuclei to constantly invert their polarization, averaging out the effect of this coupling on the observed nuclei.
High resolution NMR spectrometers used for chemical analysis are commonly equipped with pulsed-field-gradient coils, the two or more coils required for heteronuclear decoupling, and are capable of recording the signal induced by precessing $^1$H nuclei with extremely high accuracy and temporal resolution: making them ideal for this kind of tomographic measurement.

\subsection{Simultaneous gradient and spin echo with composite pulses ---}
\begin{figure}
\centerline{\includegraphics[width=15.5cm]{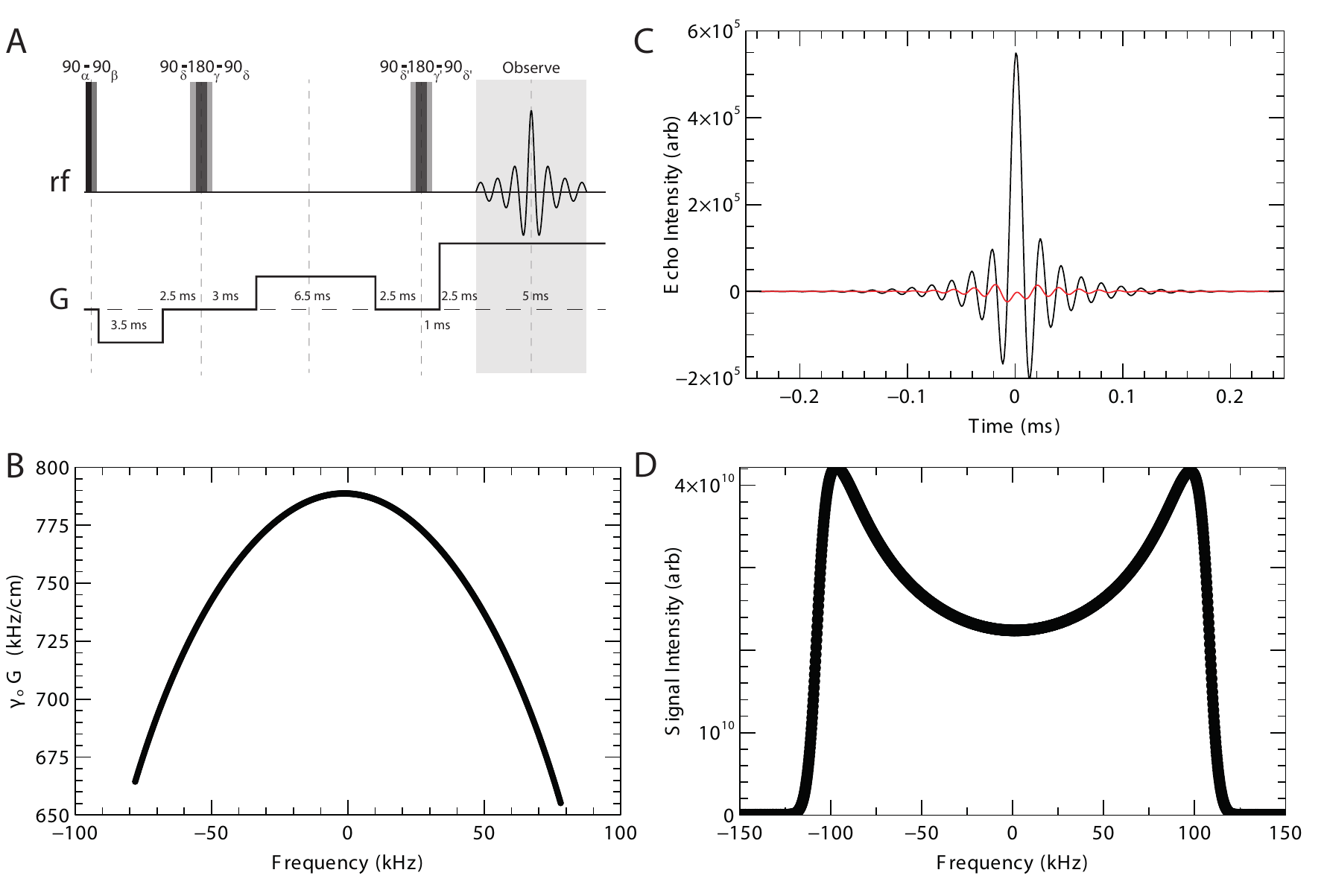}}
\caption{\textbf{Spin and gradient echo pulse sequence for $^1$H imaging and gradient calibration}
(A) Gradient echo sequence for proton densitometry.  Double spin-echo with composite excitation and inversion pulses corrects for static field gradients and probe power deficiencies.
(B) Magnitude of field gradient applied by PFG coil measured by diffusion attenuation.
(C) Real (black) and imaginary (red) components of the echo signal produced by a uniform D$_2$O/H$_2$O sample. (D) Absolute value of the Fourier transform of echo in (C).}
\label{fig:Hseq}
\end{figure}

The simple rf and field gradient pulse sequence described in Fig.~\ref{fig:seqschem} is adequate for measurements where qualitatively correct densities are sufficient.
To improve the accuracy of the measurement of proton densities we use the pulse sequence depicted in Fig.~\ref{fig:Hseq}A.
To the simple bipolar gradient echo we add a two composite spin refocusing pulses.
These refocusing pulses minimize the loss of signal that may arise from lateral variations in $B_0$ and the slight inhomogeneities in the magnetic field caused by the difference in the magnetic susceptibilities of the oil and water.
We chose a composite 90$_\alpha$-90$_\beta$ excitation pulse paired with matched composite 90$_\delta$-180$_\gamma$-90$_\delta$ refocusing pulses to compensate for the non-uniformity of the rf field generated by the probe coil.
The refocusing pulses must be matched to avoid introducing spurious phase offsets \cite{levitt_composite_1986,hurlimann_carrpurcell_2001,levitt_compensation_1981}.
We account for the finite time of the excitation pulse by shortening the time before the first refocusing pulse by  $\frac{2\,\tau_{90}}{\pi}$, where $\tau_{90}$ is the duration of a single excitation pulse \cite{hurlimann_optimization_2001}.
We compensate for imperfections in the shape of $G(t)$ caused by the finite slew rate of the amplifier by adjusting the start of the read gradient so that the echo is centered,
and place the first gradient pulse immediately after the composite excitation pulse to minimize the effects of radiation damping \cite{krishnan_radiation_2013}.
This last artifact is due to the high concentration of spins in the measurement volume coupled to a probe coil with a high resonant quality factor ($Q$).
When all the spins are excited at once, the precessing spins induce a current in the rf coil that then produces an oscillating magnetic field that interacts with the spins.
When the proton density or the quality factor of the coil are low, this second order effect is negligible: neither condition is true of this measurement.
By applying a large gradient immediately after the spins are excited we spread out their resonant frequency, which quickly leads to destructive interference that is not reversed until the read gradient refocuses the signal.
The full phase cycling table is reported in Table~\ref{tab:megaphasetable}.

\begin{table}[h]
\centering
\begin{tabular}{|c|cccccc||c|cccccc|}
\hline
  & \multicolumn{6}{c||}{Rotation Axis} & & \multicolumn{6}{c|}{Rotation Axis}\\
\cline{2-7}  \cline{9-14}
Cycle & $\alpha$ & $\beta$ & $\gamma$ & $\delta$ & $\gamma^\prime$ & $\delta^\prime$ & Cycle & $\alpha$ & $\beta$ & $\gamma$ & $\delta$ & $\gamma^\prime$ & $\delta^\prime$\\
	\hline
1 & y & x & x & y & x & y & 17 & y & x & x & -y & x & -y \\
2 & y & -x & x & y & x & y & 18 & y & -x & x & -y & x & -y \\
3 & y & x & -x & -y & x & y & 19 & y & x & -x & y & x & -y \\
4 & y & -x & -x & -y & x & y & 20 & y & -x & -x & y & x & -y \\
5 & y & x & x & y & -x & -y & 21 & y & x & x & -y & -x & y \\
6 & y & -x & x & y & -x & -y & 22 & y & -x & x & -y & -x & y \\
7 & y & x & -x & -y & -x & -y & 23 & y & x & -x & y & -x & y \\
8 & y & -x & -x & -y & -x & -y & 24 & y & -x & -x & y & -x & y \\
9 & y & x & -y & x & -y & x & 25 & y & x & -y & -x & -y & -x \\
10 & y & -x & -y & x & -y & x & 26 & y & -x & -y & -x & -y & -x \\
11 & y & x & y & -x & -y & x & 27 & y & x & y & x & -y & -x \\
12 & y & -x & y & -x & -y & x & 28 & y & -x & y & x & -y & -x \\
13 & y & x & -y & x & y & -x & 29 & y & x & -y & -x & y & x \\
14 & y & -x & -y & x & y & -x & 30 & y & -x & -y & -x & y & x \\
15 & y & x & y & -x & y & -x & 31 & y & x & y & x & y & x \\
16 & y & -x & y & -x & y & -x & 32 & y & -x & y & x & y & x \\
\hline
\end{tabular}
\caption{Partial phase cycling table for the compensated, proton imaging pulse sequence.  Phases of the rotation pulses are specified by the axis about which they rotate the magnetization.  Full phase cycling table includes the reflected sequence ($\alpha\to-\alpha$, $\beta\to-\beta$, $\gamma\to-\gamma$, etc.).}
\label{tab:megaphasetable}
\end{table}

\subsection{Calibration and Data Analysis}

We test this sequence with a 5\,mm NMR tube (542-PP-7, Wilmad-LabGlass) filled with a 40:60 weight ratio mix of H$_2$O (Millipore) and D$_2$O (99.9\,\% DLM-4, Cambridge Isotope).
We load the tube in the magnet, set the temperature of the forced air supply to 23$^{\circ}$C, and wait for thermal gradients to dissipate.
We use an inversion recovery pulse sequence, a population inverting 180$^{\circ}$ pulse followed by a variable waiting period and a 90$^{\circ}$ excitation pulse, to measure the longitudinal relaxation time of the spins, T$_1\!\approx$\,4\,s.
This relaxation time sets the rate at which the memory of past measurements is erased.
We choose to wait 20\,s between measurements for the magnetization of the spins to return to equilibrium.
We show a representative echo gathered using this sequence and the absolute value of its Fourier transform in Fig.~\ref{fig:Hseq}C-D.

If the magnetic field gradient and the signal intensity generated by the rf coil were perfectly uniform, it would be possible to directly convert the Fourier transform of the signal intensity, $|I(\omega)|$, into a measurement of the density profile, $\rho(z)$; however, the rf and gradient coils are not infinitely long and have subtle imperfections that affect the measured signal intensity.
For materials with small magnetic susceptibilities, the variations in the field gradient and detector response can be measured on a reference sample and removed.

\subsubsection{Magnetic field gradient calibration}

The magnetic field experienced by nuclear spins attached to diffusing molecules will vary depending on their position.
Consequently, the accumulated phase will depend on the molecule's trajectory:
\begin{equation*}
	\psi(\vec{r},t)=\gamma_0\int_{0}^T \!(B(z(t),t)-B_0)\,\mathrm{d}t
\end{equation*}
and perfect refocusing is not generally possible.
When averaged over a macroscopic number of spins that follow unbiased random walks, this dephasing becomes an incoherent attenuation of the raw signal \cite{price_pulsed-field_1998,price_pulsed-field_1997}.
For unconstrained diffusion, the attenuation of the signal can be expressed as:
\begin{equation*}
	\log\frac{I(G(\omega))}{I(G(\omega)=0)}=-A\,D\,\gamma_0^2\, G(\omega)^2
\end{equation*}
where $D$ is the diffusion constant of the molecules, and $A$ is a constant determined by the timing of the pulse sequence.
The diffusion constants for H$_2$O/D$_2$O water mixtures are know \cite{holz_temperature-dependent_2000}, and $A$ can be computed easily.
We thus compute $\frac{\mathrm{d}\omega}{\mathrm{d}z}\!=\!\gamma_0\,G(\omega)$ from the signal attenuation measured at every frequency and recover $z(\omega)$.
The gradient intensity profile measured using our H$_2$O/D$_2$O reference is shown in Fig.~\ref{fig:Hseq}B.

\subsubsection{rf signal intensity calibration}

A resonant rf field rotates the magnetization about an axis normal to $B_0\hat{z}$ by an angle that is proportional to the rf pulse duration and field intensity.
rf coils for high resolution NMR are designed to have a nearly binary induced field profile:
The center of the coil produces a strong, uniform field that excites a large population spins evenly, while the field intensity towards the edges of the coil decays rapidly to minimize signal from spins that receive less than a full dose of the applied rf pulse.
The same coil that produces these rf pulses is also used to measure the current induced by the precessing magnetization: the excitation and collection efficiencies are, therefore, equal.
These variations in the spatial response of the probe are reflected in the rapid decay of the signal measured at higher frequencies (Fig.~\ref{fig:Hseq}D).
To minimize the error due to miscalibrated pulse power, we choose to keep data from regions of the spectrum that are far enough from the edges of the coil.
We also compensate for smaller variations in the uniformity of the rf field intensity with composite excitation and inversion pulses~\cite{levitt_composite_1986,hurlimann_carrpurcell_2001, levitt_compensation_1981}.

Despite these corrections, it is still not possible to find a single rf pulse duration that maximizes the measured signal intensity over the entire middle portion of the coil.
The rf field induced by the coil in our spectrometer is slightly larger near the top of the sample than the bottom.
The duration of the pulse that optimizes the signal collected from the top of our active region is close to 8\% shorter than that which maximizes the signal from the bottom.
Though our sequence reduces the difference between the optimal and median signal intensity to $\lesssim$\,1\%, this systematic error is unacceptable.
We address this shortcoming by recording echos using pulse lengths that bracket the median pulse duration by at least $\pm$\,5\%, fit polynomials to the intensity measured at every frequency, and find the optimal pulse duration, $\tau(\omega)$, and the optimal signal intensity, $I^*(\omega)$,  for each frequency.
For samples with proton concentrations of at least 10\,M we find that the reproducibility of this measurement is $\approx \pm$\,0.03\%.

\subsubsection{Diffusion attenuation calibration}

The free diffusion of molecules in a field gradient provides a convenient way to calibrate the field profile and generates a uniform attenuation that can be divided out and ignored in uniform samples.
However, the diffusion of molecules contained inside of droplets or in the space between droplets is hindered by these impenetrable interfaces~\cite{sen_time-dependent_2004, callaghan_principles_1993, zielinski_combined_2003, zielinski_relaxation_2000, zielinski_effects_2003}.
Moreover, small differences in the magnetic susceptibilities of the materials in the imaging volume produce static magnetic field gradients that also contribute to the attenuation of the signal.
The attenuation of the magnetic signal in non-uniform materials thus depends on the local structure of the medium in a way that is not trivial.

When the static field gradients induced by the contrast in magnetic susceptibilities is small, their contribution to the signal attenuation can be nearly eliminated using the 180$^{\circ}$ refocusing pulses included in our sequence; the remaining signal attenuation then depends on $|G^2|$.
We record echos at several values of $G$ and compute the $G\!\to$\,0 limit of $I^*(\omega)$.
The resulting signal is reproducible and well defined, but is still affected by the shape of the spatial response function of the rf coil, and possible variations in the shape of the tubes that hold the sample.
However, these contributions do not depend on the material contained in the tube, so we generate a reference for each tube by performing the full set of measurements required to obtain $I^*(\omega,\, G\!\to$\,0$)$ on uniform fluid samples.
When the pulse sequence duration is much shorter than the transverse decorrelation time of the magnetization, $T_{2}^*$, and the time between measurements is much longer that the longitudinal relaxation time, $T_1$, the density profile is recovered by dividing the measured $I^*(\omega,\, G\!\to$\,0$)$ by the stored reference.

\subsection{Proton Imaging ---}
We load a 45\,vol.\,\% solution of tetraglyme in D$_2$O into a 5\,mm NMR tube and collect a reference measurement of the coil response and tube shape.
We then clean the tube by rinsing it several times with a 2\,mM solution of SDBS in D$_2$O, add enough of a 20\,vol.\,\% dispersion of the PDMS/TCE droplets described in section 1.5 to it for the resulting sediment to nearly fill the imaging volume, and allow the droplets to settle while the tube is held vertically.
After the sediment is formed we add more of the SDBS-D$_2$O solution to the tube and seal it with fluorinated grease (Krytox LVP, Dupont) and a Teflon cap (WG-1264-5, Wilmad-LabGlass).
We anneal the sediment by storing the tube alternatively at 20$^{\circ}$C and 27$^{\circ}$C, changing the temperature eight times over two days.
The temperature cycling causes droplet volumes to fluctuate by $\sim$1\%, but the volume of the borosilicate glass tube remains almost constant.
This mismatch converts the changes in droplet volume into uniaxial strains.

After waiting at least one week for the sample to equilibrate at room temperature we load it into the spectrometer and set the temperature control to 23$^{\circ}$C.
While we wait for the temperature of the sample to equilibrate we measure the transverse, $T_2^*$, and longitudinal relaxation times, $T_1\!\approx$\,1.7\,s, and shim the magnetic field, $B_0$.
We set the time between measurements to 12\,s to eliminate any coupling between successive measurements.
We monitor the position of the interface while all the necessary combinations of rf pulse length and field gradient strength samplings are executed.
This procedure lasts about five hours, and we use the shape of sequential echos to confirm that the interface has not move appreciably during this time.

\subsection{Internal Deuterium Reference ---}

When the reference sample is not composed of the same material that composes the sample we want to image, the measured signal is proportional to the mass density, but is not equal to it.
We chose to use a D$_2$O/tetraglyme solution with a similar proton concentration as a reference for this sample because of its chemical homogeneity and favorable diffusion constant;
however, without a reference that matches the signal of the pure PDMS/TCE mixture exactly, we cannot produce quantitative volume fraction measurements from proton signals alone.
Instead, we image the deuterium in the surrounding fluid and use this as an absolute, internal reference of the volume fraction, $\phi$.

\begin{figure}
\centerline{\includegraphics[width=15.5cm]{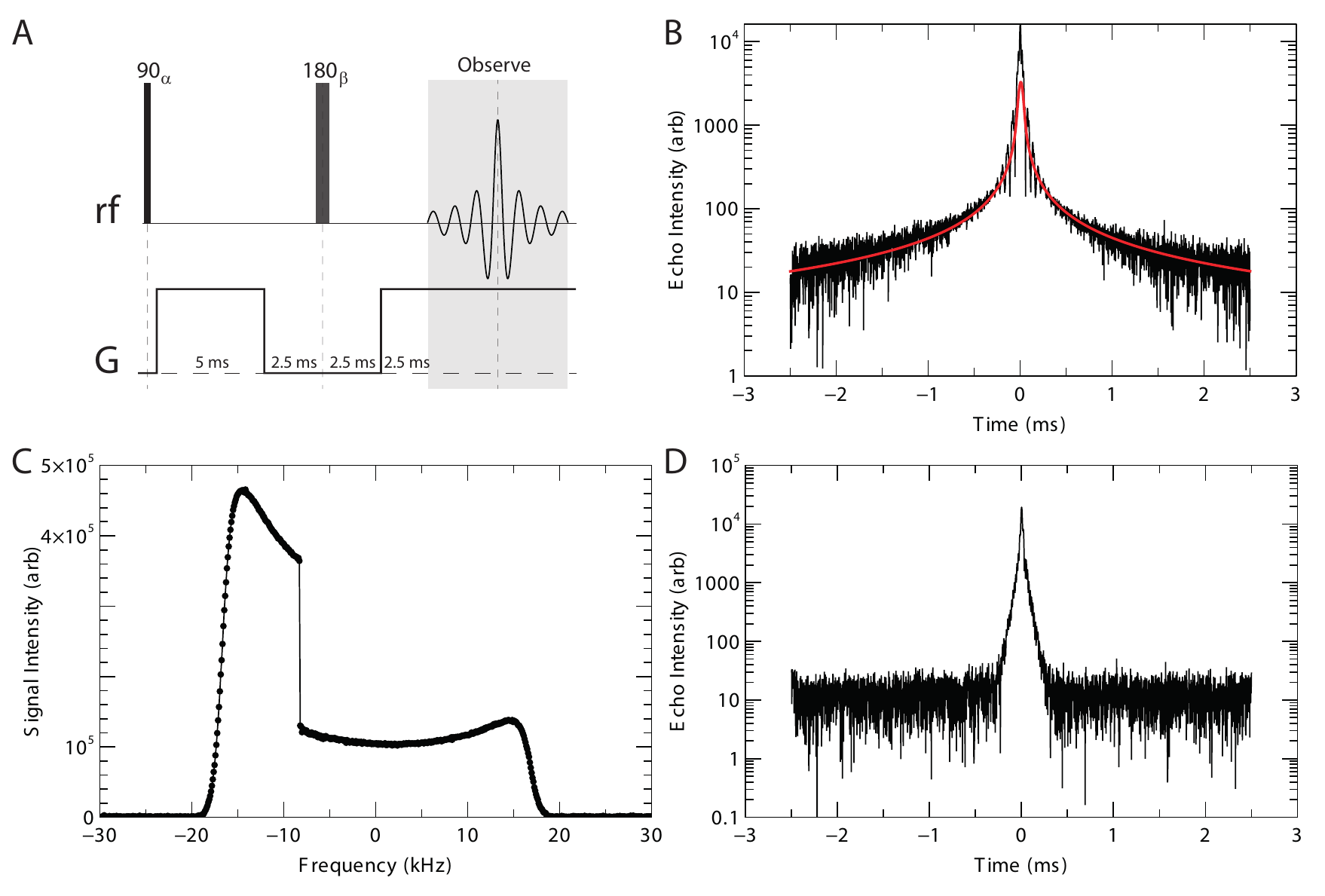}}
\caption{\textbf{Gradient echo sequence for deuterium densitometry and echo ringing correction}
(A) Gradient echo sequence for deuterium densitometry.  Single spin-echo reduces the effects of static field gradients.
(B) Absolute value of the measured echo (black) and Fourier transform of the step-Gaussian fit (red) for deuterium imaging of sedimented droplets.
(C) Fourier transform of the extended echo signal produced by subtracting and adding back a step-Gaussian density from the measured echo.
(D) Absolute value of the difference between the measured echo and the step-Gaussian function.}
\label{fig:Dseq}
\end{figure}

The deuterium nucleus has a spin of 1, and the three angular momentum states are coupled by its electric quadrupole moment.
This coupling can make deuterium spectroscopy more complicated than proton spectroscopy, but, when nuclei are free to rotate quickly, this coupling only results in faster relaxation rates.
The resonant frequency of deuterium is also much smaller than that of hydrogen, which reduces the signal intensity and extends the precession times by an order of magnitude~\cite{callaghan_principles_1993, callaghan_measurement_1983, dvinskikh_anisotropic_2002, holz_calibration_1991}.
To compensate for these effects we use larger gradient strengths, and shorter, simpler pulse sequences (Fig.~\ref{fig:Dseq}A and Table~\ref{tab:2Hphasetable}).
This makes it difficult to obtain high resolution density profiles from deuterium imaging alone;
however, all we need to calibrate the proton signal intensity is an accurate measurement of the ratio of the deuterium signal just above the top of the sedimented emulsion to that just below it: which provides an internal reference for $\phi_c$, the critical volume fraction.
We still need to compute the $G\!\to$\,0 limit of the signal intensity to measure this number, but thorough pulse timing calibrations are not necessary because the measurement only needs to be accurate in a small neighborhood of the top of the sediment.
A representative echo signal is shown in Fig.~\ref{fig:Dseq}B.

\begin{table}
\centering
\begin{tabular}{|c|ccc||c|ccc|}
\hline
  & \multicolumn{3}{c|}{Rotation Axis}  & & \multicolumn{3}{c|}{Rotation Axis}\\
\cline{2-4} \cline{6-8} 
Cycle & & $\alpha$ & $\beta$ & Cycle & & $\alpha$ & $\beta$\\
	\hline
1 & & y & y & 9 & & -x & -x\\
2 & & y & -y & 10 & & -x & x\\
3 & & y & x & 11 & & -x & y\\
4 & & y & -x & 12 & & -x & -y\\
5 & & -y & -y & 13 & & x & x\\
6 & & -y & y & 14 & & x & -x\\
7 & & -y & -x & 15 & & x & -y\\
8 & & -y & x & 16 & & x & y\\
\hline
\end{tabular}
\caption{Complete phase cycling table for the deuterium imaging pulse sequence.  Phases of the rotation pulses are specified by the axis about which they rotate the magnetization.}
\label{tab:2Hphasetable}
\end{table}

The sharp interface between the emulsion and supernatant causes a slowly-decaying ringing in the echo signal, and, since the precession frequency is much smaller than that of hydrogen, the measurement time required to fully capture this slowly decaying oscillation becomes comparable to the transverse decorrelation time of the precessing magnetization.
However, because the ringing is caused by a discontinuous density profile, we can fit this oscillation to one produced by a model step discontinuity and subtract it from the echo.
We choose a step with the form:
\begin{equation*}
	\hat{\rho}(\omega)=b\cdot(1 + c\cdot(\omega-\omega^*))H(\omega-\omega^*)\,e^{-(\omega-\omega_0)^2/a^2}e^{i(\theta_0+\theta_1\cdot(\omega-\omega^*))}
\end{equation*}
where $H(\cdot)$ is the Heaviside step function, $\omega^*$ is the position of the interface, $\theta_0$ and $\theta_1$ adjust the phase and timing of the echo,  $b$ sets the magnitude of the jump in the density, $c$ corrects for higher order oscillations due to discontinuities in the slope of the signal across the discontinuity, and $a$ sets the width of a Gaussian cut-off.
The Fourier transform of this discontinuous function precisely captures the shape of the oscillations, and the difference between the measured echo and the model echo decays below the noise floor well within the measurement window (Fig.~\ref{fig:Dseq}D).
To recover the correct shape of the raw density profile we add $\hat{\rho}(\omega)$ back to the Fourier transform of the subtracted echo (Fig.~\ref{fig:Dseq}C).
This correction makes it possible to reduce our measurement window enough to avoid issues related to the natural decay of transverse magnetization
After repeating this process for several values of $G$, we recover the absolute value of the droplet volume fraction at the top of the sediment from the $G\!\to$\,0 limit of $\phi_c\equiv1-\frac{I^*(\omega\to\omega_{0}\,^+)}{I^*(\omega\to\omega_{0}\,^-)}$.
We use this value to normalize the density profile presented in Fig.~4B of the main text.


